\documentclass[letterpaper,twocolumn,10pt]{article}
\usepackage{usenix2019_v3}

\usepackage{tikz}
\usetikzlibrary{decorations.pathreplacing}
\usepackage{amsmath}
\usepackage{amsthm}
\usepackage{algorithm}  
\usepackage{algorithmic}
\usepackage{color}
\usepackage{colortbl,booktabs}
\usepackage{multirow}
\usepackage{subfigure}
\usepackage{color,xcolor}
\newtheorem{myDef}{Definition}
\newtheorem{theorem}{Theorem}
\newcommand{\tabincell}[2]{
\begin{tabular}{@{}#1@{}}#2\end{tabular}
}
\usepackage{filecontents}

\begin{document}
%-------------------------------------------------------------------------------

%don't want date printed
\date{}

\title{\Large \bf {Secure Watermark for Deep Neural Networks with Multi-task Learning}}

%for single author (just remove % characters)
\author{
{\rm Fangqi Li, Shilin Wang}\\
School of Electronic Information and Electrical Engineering,\\
Shanghai Jiao Tong University,\\
\texttt{ \{solour\_lfq,wsl\}@sjtu.edu.cn}
% copy the following lines to add more authors
% \and
% {\rm Name}\\
%Name Institution
} % end author

\maketitle

%-------------------------------------------------------------------------------
\begin{abstract}
%-------------------------------------------------------------------------------
Deep neural networks are playing an important role in many real-life applications.
After being trained with abundant data and computing resources, a deep neural network model providing service is endowed with economic value.
An important prerequisite in commercializing and protecting deep neural networks is the reliable identification of their genuine author.
To meet this goal, watermarking schemes that embed the author's identity information into the networks have been proposed.
However, current schemes can hardly meet all the necessary requirements for securely proving the authorship and mostly focus on models for classification. 
To explicitly meet the formal definitions of the security requirements and increase the applicability of deep neural network watermarking schemes, we propose a new framework based on multi-task learning.
By treating the watermark embedding as an extra task, most of the security requirements are explicitly formulated and met with well-designed regularizers, the rest is guaranteed by using components from cryptography.
Moreover, a decentralized verification protocol is proposed to standardize the ownership verification.
The experiment results show that the proposed scheme is flexible, secure, and robust, hence a promising candidate in deep learning model protection. 
\end{abstract}

%-------------------------------------------------------------------------------
\section{Introduction}
%-------------------------------------------------------------------------------
Deep neural network (DNN) is spearheading artificial intelligence with broad application in assorted fields including computer vision~\cite{cv1,cv2,cv3}, natural language processing~\cite{nlp1,nlp2,nlp3}, internet of things~\cite{iot1,iot2,iot3}, etc.
Increasing computing resources and improved algorithms have boosted DNN as a trustworthy agent that outperforms humans in many disciplines.

To train a DNN is much more expensive than to use it for inference.
A large amount of data has to be collected, preprocessed, and fed into the model.
Following the data preparation is designing the regularizers, tuning the (hyper)parameters, and optimizing the DNN structure.
Each round of tuning involves thousands of epochs of backpropagation, whose cost is about 0.005\$ averagely regarding electricity consumption.\footnote{Assume that one-kilowatt-hour electricity costs 0.1\$. One epoch for training a DNN can consume over three minutes on four GPUs, each functions at around 300W.}
On the contrary, using a published DNN is easy, a user simply propagates the input forward.
Such an imbalance between DNN production and deployment calls for recognizing DNN models as intellectual properties and designing better mechanisms for authorship identification against piracy.

DNN models, as other multi-media objects, are usually transmitted in public channels.
Hence the most influential methods for protecting DNNs as intellectual properties is digital watermark~\cite{zhang2018protecting}.
To prove the possession of an image, a piece of music, or a video, the owner resorts to a watermarking method that encodes its identity information into the media.
After compression, transmission, and slight distortion, a decoder should be able to recognize the identity from the carrier~\cite{9100612}.

As for DNN watermarking, researchers have been following a similar line of reasoning~\cite{uchida2017embedding}.
In this paper, we use \emph{host} to denote the genuine author of a DNN model. 
The \emph{adversary} is one who steals and publishes the model as if it is the host.
To add watermarks to a DNN, some information is embedded into the network along with the normal training data.
After adversaries manage to steal the model and pretend to have built it on themselves, a verification process reveals the hidden information in the DNN to identify the authentic host.
In the DNN setting, watermark as additional security insurance should not sacrifice the model's performance.
This is called the \emph{functionality-preserving} property.
Meanwhile, the watermark should be robust against the adversaries' modifications to the model.
Many users fine-tune (FT) the downloaded model on a smaller data set to fit their tasks.
In cases where the computational resource is restricted (especially in the internet of things), a user is expected to conduct neuron pruning (NP) to save energy.
A prudent user can conduct fine-pruning (FP)~\cite{liu2018fine} to eliminate potential backdoors that have been inserted into the model. 
These basic requirements, together with other concerns for integrity, privacy, etc, make DNN watermark a challenge for both machine learning and security communities.

The diversity of current watermarking schemes originates from assumptions on whether or not the host or the notary has white-box access to the stolen model. 

If the adversary has stolen the model and only provided an API as a service then the host has only black-box access to the possibly stolen model.
In this case, the \emph{backdoor-based watermarking schemes} are preferred.
A DNN with a backdoor yields special outputs on specific inputs.
For example, it is possible to train an image classification DNN to classify all images with a triangle stamp on the upper-left corner as cats.
Backdoor-based watermark was pioneered by~\cite{zhang2018protecting}, where a collection of images is selected as the \emph{trigger set} to actuate misclassifications.
It was indicated in~\cite{adi2018turning,zhu2020secure} that cryptological protocols can be used with the backdoor-based watermark to prove the integrity of the host's identity.
For a more decent way of generating triggers, Li et al. proposed in~\cite{li2019prove} to adopt a variational autoencoder (VAE), while Le Merrer et al. used adversarial samples as triggers~\cite{le2020adversarial}.
Li et al. proposed \texttt{Wonder Filter} that assigns some pixels to values in $[-2000,2000]$ and adopted several tricks to guarantee the robustness of watermark embedding in~\cite{li2019persistent}.
In~\cite{yao2019latent}, Yao et al. illustrated the performance of the backdoor-based watermark in transfer learning and concluded that it is better to embed information in the feature extraction layers.

The backdoor-based watermarking schemes are essentially insecure given various methods of backdoor elimination~\cite{chen2019refit,liu2020removing,li2021neural}. 
Liu et al. showed in~\cite{liu2019abs} that a heuristic and
biomorphic method can detect backdoor in a DNN.
In~\cite{shafieinejad2019robustness}, Shafieinejad et al. claimed that it is able to remove watermarks given the black-box access of the model.
Namba et al. proposed another defense using VAE against backdoor-based watermarking methods in~\cite{namba2019robust}.
Even without these specialized algorithms, model tuning such as FP~\cite{liu2018fine,tung2017fine} can efficiently block backdoor and hence the backdoor-based watermark. 

If the host can obtain all the parameters of the model, known as the white-box access, then the \emph{weight-based watermarking schemes} are in favor.
Although this assumption is strictly stronger than that for the black-box setting, its practicality remains significant. 
For example, the sponsor of a model competition can detect plagiarists that submit models slightly tuned from those of other contestants by examing the watermark.
This legitimate method is better than checking whether two models perform significantly different on a batch of data, which is still adopted by many competitions.\footnote{\url{http://host.robots.ox.ac.uk:8080/leaderboard/main_bootstrap.php}} 
As another example, the investor of a project can verify the originality of a submitted model from its watermark. 
Such verification prevents the tenderers from submitting a (modified) copy or an outdated and potentially backdoored model. 

Uchida et al. firstly revealed the feasibility of incorporating the host's identity information into the weights of a DNN in~\cite{uchida2017embedding}.
The encoding is done through a regularizer that minimizes the distance between a specific weight vector and a string encoding the author's identity.
The method in~\cite{guan2020reversible} is an attempt of embedding message into the model's weight in a reversible manner so that a trusted user can eliminate the watermark's influence and obtain the clean model.
Instead of weights, Davish et al. proposed \texttt{Deepsigns}\cite{darvish2019deepsigns} that embeds the host's identity into the statistical mean of the feature maps of a selected collection of samples, hence better protection is achieved.
%A summary of the comparison among early watermarking methods can be found in~\cite{chen2018performance}.

So far, the performance of a watermarking method is mainly measured by the decline of the watermarked model's performance on normal inputs and the decline of the identity verification accuracy against model fine-tuning and neuron pruning.
However, many of the results are empirical and lack analytic basis~\cite{uchida2017embedding,darvish2019deepsigns}.
Most watermarking methods are only designed and examined for DNNs for image classification, whose backdoors can be generated easily. 
This fact challenges the universality of adopting DNN watermark for practical use.
Moreover, some basic security requirements against adversarial attacks have been overlooked by most existing watermarking schemes.
For example, the method in~\cite{zhang2018protecting} can detect the piracy, but it cannot \emph{prove} to any third-party that the model belongs to the host.
As indicated by Auguste Kerckhoff's principle~\cite{knoll2018adapting}, the security of the system should rely on the secret key rather than the secrecy of the algorithm.
Methods in~\cite{zhang2018protecting, darvish2019deepsigns, uchida2017embedding} are insecure in this sense since an adversary knowing the watermark algorithm can effortlessly claim the authorship.
The influence of watermark overwriting is only discussed in~\cite{adi2018turning,li2019persistent,darvish2019deepsigns}.
The security against ownership piracy is only studied in~\cite{zhu2020secure,li2019persistent,guan2020reversible}.

In order to overcome these difficulties, we propose a new white-box watermarking model for DNN based on multi-task learning (MTL)~\cite{mtl, sener2018multi, kendall2018multi}.
By turning the watermark embedding into an extra task, most security requirements can be satisfied with well-designed regularizers.
This extra task has a classifier independent from the backend of the original model, hence it can verify the ownership of models designed for tasks other than classification. 
Cryptological protocols are adopted to instantiate the watermarking task, making the proposed scheme more secure against watermark detection and ownership piracy.
To ensure the integrity of authorship identification, a decentralized verification protocol is designed to authorize the time stamp of the ownership and invalid the watermark overwriting attack.
The major contributions of our work are three-fold:
\begin{enumerate}
\item We examine the security requirements for DNN watermark in a comprehensive and formal manner.
\item A DNN watermarking model based on MTL, together with a decentralized protocal, is proposed to meet all the security requirements.
Our proposal can be applied to DNNs for tasks other than image classification, which were the only focus of previous works.
\item Compared with several state-of-the-art watermarking schemes, the proposed method is more robust and secure. 
\end{enumerate}

\section{Threat Model and Security Requirements}
\label{section:2}
It is reasonable to assume that the adversary possesses fewer resources than the host, e.g., the entire training data set is not exposed to the adversary, and/or the adversary's computation resources are limited.
Otherwise, it is unnecessary for the adversary to steal the model.
Moreover, we assume that the adversary can only tune the model by methods such as FT, NP or FP.
Such modifications are common attacks since the training code is usually published along with the trained model. 
Meanwhile, such tuning is effective against systems that only use the hash of the model as the verification. 
On the other hand, it is hard and much involved to modify the internal computational graph of a model. 
It is harder to adopt model extraction or distillation that demands much data and computation~\cite{kesarwani2018model,polino2018model}, yet risks performance and the ability of generalization. 
Assume that the DNN model $M$ is designed to fulfil a \emph{primary task}, $\mathcal{T}_{\text{primary}}$, with dataset $\mathcal{D}_{\text{primary}}$, data space $\mathcal{X}$, label space $\mathcal{Y}$ and a metric $d$ on $\mathcal{Y}$.

\subsection{Threat Model}
We consider five major threats to the DNN watermarks.

\subsubsection{Model tuning}
An adversary can tune $M$ by methods including: (1) FT: running backpropagation on a local dataset, (2) NP: cut out links in $M$ that are less important, and (3) FP: pruning unnecessary neurons in $M$ and fine-tuning $M$. 
The adversary's local dataset is usually much smaller than the original training dataset for $M$ and fewer epochs are needed.
FT and NP can compromise watermarking methods that encode information into $M$'s weight in a reversible way~\cite{guan2020reversible}.
Meanwhile,~\cite{liu2018fine} suggested that FP can efficiently eliminate backdoors from image classification models and watermarks within.

\subsubsection{Watermark detection}
If the adversary can distinguish a watermarked model from a clean one, then the watermark is of less use since the adversary can use the clean models and escape copyright regulation.
The adversary can adopt backdoor screening methods~\cite{yang2019neural,wang2019neural,8835365} or reverse engineering~\cite{hua2018reverse,batina2019csi} to detect and possibly eliminate backdoor-based watermarks.
For weight-based watermarks, the host has to ensure that the weights of a watermarked model do not deviate from that of a clean model too much. Otherwise, the property inference attack~\cite{ganju2018property} can distinguish two models.

\subsubsection{Privacy concerns}
As an extension to detection, we consider an adversary who is capable of identifying the host of a model without its permission as a threat to privacy. 
A watermarked DNN should expose no information about its host unless the host wants to. 
Otherwise, it is possible that models be evaluated not by their performance but by their authors. 

\subsubsection{Watermark overwriting}
Having obtained the model and the watermarking method, the adversary can embed its watermark into the model and declare the ownership afterward.
Embedding an extra watermark only requires the redundancy of parameter representation in the model.
Therefore new watermarks can always be embedded unless one proves that such redundancy has been depleted, which is generally impossible.
A concrete requirement is: the insertion of a new watermark should not erase the previous watermarks.

For a model with multiple watermarks, it is necessary that an an incontrovertible time-stamp is included into ownership verification to break this \emph{redeclaration dilemma}.

\subsubsection{Ownership piracy}
Even without tuning the parameters, model theft is still possible.
Similar to~\cite{li2019prove}, we define \emph{ownership piracy} as attacks by which the adversary claims ownership over a DNN model without tuning its parameters or training extra learning modules.
For zero-bit watermarking schemes (no secret key is involved, the security depends on the secrecy of the algorithm), the adversary can claim ownership by publishing a copy of the scheme.
For a backdoor-based watermarking scheme that is not carefully designed, the adversary can detect the backdoor and claim that the backdoor as its watermark.

The secure watermarking schemes usually make use of cryptological protocols\cite{zhu2020secure,li2019persistent}.
In these schemes, the adversary is almost impossible to pretend to be the host using any probabilistic machine that terminates within time complexity polynomial to the security parameters (PPT). 

\subsection{Formulating the Watermarking Scheme}

We define a watermarking scheme with security parameters $N$ as a probabilistic algorithm $\texttt{WM}$ that maps $\mathcal{T}_{\text{primary}}$ (the description of the task, together with the training dataset $\mathcal{D}_{\text{primary}}$), a description of the structure of the DNN model $\mathcal{M}$ and a secret key denoted by $\texttt{key}$ to a pair $\left(M_{\text{WM}},\texttt{verify}\right)$:
$$\texttt{WM}:\left(M_{\text{WM}},\texttt{verify}\right)\leftarrow \left(N,\mathcal{T}_{\text{primary}},\mathcal{M},\texttt{key}\right),$$
where $M_{\text{WM}}$ is the watermarked DNN model and $\texttt{verify}$ is a probabilistic algorithm with binary output for verifying ownership.
To verify the ownership, the host provides $\texttt{verify}$ and $\texttt{key}$. 
A watermarking scheme should satisfy the following basic requirements for \emph{correctness}:
\begin{equation}
\label{equation:1}
\text{Pr}\left\{\texttt{verify}(M_{\text{WM}},\texttt{key})=1 \right\}\geq 1-\epsilon,
\end{equation}
\begin{equation}
\label{equation:2}
\text{Pr}_{
\begin{scriptsize}
\begin{aligned}
&M'\text{ irrelevent to } M_{\text{WM}}\text{,} \\
&\text{or }\texttt{key}'\neq \texttt{key}
\end{aligned}
\end{scriptsize}
}
\left\{\texttt{verify}(M',\texttt{key}')=0 \right\}\geq 1-\epsilon,
\end{equation}
where $\epsilon\in(0,1)$ reflects the security level.
Condition \eqref{equation:1} suggests that the verifier should always correctly identify the authorship while \eqref{equation:2} suggests that it only accepts the correct key as the proof and it should not mistake irrelevant models as the host's.

The original model trained without being watermarked is denoted by $M_{\text{clean}}$.
Some researchers~\cite{guan2020reversible} define $\texttt{WM}$ as a mapping from $\left(N,M_{\text{clean}},\texttt{key}\right)$ to $\left(M_{\text{WM}},\texttt{verify}\right)$, which is a subclass of our definition.

\subsection{Security Requirements}
Having examined the toolkit of the adversary, we formally define the security requirements for a watermarking scheme.

\subsubsection{Functionality-preserving}
The watermarked model should perform slightly worse than, if not as well as, the clean model.
The definition for this property is:
\begin{equation}
\label{equation:4}
\text{Pr}_{(x,y)\sim \mathcal{T}_{\text{primary}}}\left\{d(M_{\text{clean}}(x),M_{\text{WM}}(x))\leq \delta\right\} \geq 1-\epsilon,
\end{equation}
which can be examined \emph{a posteriori}.
However, it is hard to explicitly incorporate this definition into the watermarking scheme.
Instead, we resort to the following definition:
\begin{equation}
\label{equation:3}
\forall x\in\mathcal{X}\text{, }d(M_{\text{clean}}(x),M_{\text{WM}}(x))\leq \delta.
\end{equation}
Although it is stronger than~\eqref{equation:4},~\eqref{equation:3} is a tractable definition. 
We only have to ensure that the parameters of $M_{\text{WM}}$ does not deviate from those of $M_{\text{clean}}$ too much.

\subsubsection{Security against tuning}
After being tuned with the adversary's dataset $\mathcal{D}_{\text{adversary}}$, the model's parameters shift and the verification accuracy of the watermark might decline.
Let
$M'\xleftarrow[\mathcal{D}_{\text{adversary}}]{\text{tuning}} M_{\text{WM}}$
denotes a model $M'$ obtained by tuning $M_{\text{WM}}$ with $\mathcal{D}_{\text{adversary}}$.
A watermarking scheme is secure against tuning iff:
\begin{equation}
\label{equation:6}
\text{Pr}_{
\begin{scriptsize}
\begin{aligned}
&\mathcal{D}_{\text{adversary}},\\
M'&\xleftarrow[\mathcal{D}_{\text{adversary}}]{\text{tuning}} M_{\text{WM}}
\end{aligned}
\end{scriptsize}
}
\left\{\texttt{verify}(M',\texttt{key})=1 \right\}\geq 1-\epsilon.
\end{equation}
%where the probability is computed by integrating out the randomness in selecting $\mathcal{D}_{\text{adversary}}$, generating $M'$ and running $\texttt{verify}$.
To meet \eqref{equation:6}, the host has to simulate the effects of tuning and make $\texttt{verify}(\cdot,\texttt{key})$ insensitive to them in the neighbour of $M_{\text{WM}}$.

\subsubsection{Security against watermark detection}
According to~\cite{wang2019robust}, one definition for the security against watermark detection is: no PPT can distinguish a watermarked model from a clean one with nonnegligible probability.
Although this definition is impractical due to the lack of a universal backdoor detector, it is crucial that the watermark does not differentiate a watermarked model from a clean model too much. 
Moreover, the host should be able to control the level of this difference by tuning the watermarking method.
Let $\theta$ be a parameter within $\texttt{WM}$ that regulates such difference, it is desirable that
\begin{equation}
\label{equation:detection}
M^{\infty}_{\text{WM}}=M_{\text{clean}},
\end{equation}
where $M^{\infty}_{\text{WM}}$ is the model returned from $\texttt{WM}$ with $\theta\rightarrow\infty$. 

\subsubsection{Privacy-preserving}
\label{section:2.3.4}
To protect the host's privacy, it is sufficient that any adversary cannot distinguish between two models watermarked with different $\texttt{key}$s.  
Fixing the primary task $\mathcal{T}_{\text{primary}}$ and the structure of the model $\mathcal{M}$, we first introduce an experiment $\texttt{Exp}^{\text{detect}}_{\mathcal{\mathcal{A}}}$ in which an adversary $\mathcal{A}$ tries to identify the host of a model:
\begin{algorithm}[htbp] 
\caption{$\texttt{Exp}^{\text{detect}}_{\mathcal{\mathcal{A}}}$.}  
\label{algorithm:1}  
\begin{algorithmic}[1]
\REQUIRE $N$, $\texttt{WM}$, $\texttt{key}_{0}\neq\texttt{key}_{1}$.
\STATE Randomly select $b\leftarrow \left\{0,1 \right\}$;
\STATE Generate $M_{\text{WM}}$ from $\texttt{WM}(N,\mathcal{T}_{\text{primary}},\mathcal{M},\texttt{key}_{b})$;
\STATE $\mathcal{A}$ is given $M_{\text{WM}}$, $N$, $\texttt{WM}$, $\texttt{key}_{0}$, $\texttt{key}_{1}$ and outputs $\hat{b}$.
\STATE $\mathcal{A}$ wins the experiment if $\hat{b}=b$.
\end{algorithmic}  
\label{exp:1}
\end{algorithm} 

\begin{table*}[htbp]
\caption{Security requirements and established watermarking schemes.}
\begin{center}
\begin{tabular}{c|ccccccccccc}
\toprule[1.5pt]
\textbf{Security requirement} & \tabincell{c}{Zhu. \\~\cite{zhu2020secure}.}& \tabincell{c}{Adi. \\~\cite{adi2018turning}.} & \tabincell{c}{Le Merrer.\\~\cite{le2020adversarial}.} & \tabincell{c}{Zhang.\\~\cite{zhang2018protecting}.} & \tabincell{c}{Davish. \\~\cite{darvish2019deepsigns}.} & \tabincell{c}{Li. \\~\cite{li2019persistent}.} & \tabincell{c}{Li.\\~\cite{li2019prove}.} & \tabincell{c}{Uchida. \\\cite{uchida2017embedding}.} & \tabincell{c}{Guan. \\~\cite{guan2020reversible}.} & Ours. \\
\midrule[1pt]
Functionality-preserving. &\texttt{E} & \texttt{E} & \texttt{P} & \texttt{E} & \texttt{P} & \texttt{E} & \texttt{E} & \texttt{P} & \texttt{E} & \texttt{P} \\
\midrule
Security against tuning. & \texttt{N} & \texttt{E} & \texttt{E} & \texttt{E} & \texttt{E} & \texttt{E} & \texttt{N} & \texttt{E} & \texttt{N} & \texttt{P} \\
\midrule
\tabincell{c}{Security against \\ watermark detection.} & \texttt{N} & \texttt{N} & \texttt{N} & \texttt{N} & \texttt{N} & \texttt{P} & \texttt{P} & \texttt{N} & \texttt{N} & \texttt{P} \\
\midrule
Privacy-preserving. & \texttt{N} & \texttt{N} & \texttt{N} & \texttt{N} & \texttt{N} & \texttt{N} & \texttt{N} & \texttt{N} & \texttt{N} & \texttt{P} \\
\midrule
\tabincell{c}{Security against \\ watermark overwriting.} & \texttt{N} & \texttt{E} & \texttt{N} & \texttt{N} & \texttt{E} & \texttt{E} & \texttt{N} & \texttt{N} & \texttt{N} & \texttt{E} \\
\midrule
\tabincell{c}{Security against \\ ownership piracy.} & III & II & I & I & II & III & II & I & III & III \\
\bottomrule[1pt]
\multicolumn{11}{l}{$\texttt{P}$ means the security requirement is claimed to be held by proof or proper regularizers. $\texttt{E}$ means an empirical evaluation}\\
\multicolumn{11}{l}{on the security was provided. $\texttt{N}$ means not discussion was given or insecure.}\\
\end{tabular}
\label{table:1}
\end{center}
\end{table*}

\begin{myDef}
If for all PPT adversary $\mathcal{A}$, the probability that $\mathcal{A}$ wins $\texttt{Exp}^{\text{detect}}_{\mathcal{\mathcal{A}}}$ is upper bounded by $\frac{1}{2}+\epsilon(N)$, where $\epsilon$ is a negligible function, then $\texttt{WM}$ is privacy-preserving.
\end{myDef}
The intuition behind this definition is: an adversary cannot identify the host from the model, even if the number of candidates has been reduced to two. 
Almost all backdoor-based watermarking schemes are insecure under this definition.
In order to protect privacy, it is crucial that $\texttt{WM}$ be a probabilistic algorithm and $\texttt{verify}$ depend on $\texttt{key}$. 

\subsubsection{Security against watermark overwriting}
Assume that the adversary has watermarked $M_{\text{WM}}$ with another secret key $\texttt{key}_{\texttt{adv}}$ using a subprocess of $\texttt{WM}$ and obtained $M_{\texttt{adv}}$:
$M_{\texttt{adv}}\xleftarrow[\texttt{key}_{\texttt{adv}}]{\text{overwriting}}M_{\text{WM}}$.
The overwriting watermark should not invalid the original one, formally, for any legal $\texttt{key}_{\text{adv}}$:
\begin{equation}
\label{equation:7}
\text{Pr}_{
\begin{scriptsize}
\begin{aligned}
&\texttt{key}_{\texttt{adv}},\\
&M_{\texttt{adv}}\xleftarrow[\texttt{key}_{\texttt{adv}}]{\text{overwriting}}M_{\text{WM}}
\end{aligned}
\end{scriptsize}
}\left\{\texttt{verify}(M_{\texttt{adv}},\texttt{key})=1 \right\}\geq 1-\epsilon.
\end{equation}
During which the randomness in choosing $\texttt{key}_{\texttt{adv}}$, generating $M_{\texttt{adv}}$, and computing $\texttt{verify}$ is integrated out.
A watermarking scheme meets \eqref{equation:7} is defined to be secure against watermark overwriting.
This property is usually examined empirically in the literature~\cite{adi2018turning,darvish2019deepsigns,li2019persistent}.

\subsubsection{Security against ownership piracy}
In an ownership piracy attack, the adversary pirate a model by recovering $\texttt{key}$ and forging $\texttt{verify}$ through querying $M_{\text{WM}}$ (or $\texttt{verify}$ if available).
We define three levels of security according to the efforts needed to pirate a model.
\begin{enumerate}
\item Level I: The adversary only needs to wrap $M_{\text{WM}}$ or query it for a constant number of times.
All zero-bit watermarking schemes belong to this level.
\item Level II: The adversary has to query $M_{\text{WM}}$ for a number of times that is a polynomial function of the security parameter.
The more the adversary queries, the more likely it is going to succeed in pretending to be the host.
The $\texttt{key}$ and $\texttt{verify}$, in this case, is generally simple. 
For example,~\cite{darvish2019deepsigns,adi2018turning} are of this level of security.
\item Level III: The adversary is almost impossible to pirate ownership of the model given queries of times that is a polynomial function of the security parameter.
Such schemes usually borrow methods from cryptography to generate the pseudorandomness.
Methods in~\cite{li2019persistent,zhu2020secure} are examples of this level.
\end{enumerate}
Watermarking schemes of level I and II can be adopted as theft detectors.
But the host can hardly adopt a level I/II scheme to convince a third-party about ownership. 
Using a watermarking scheme of level III, a host can prove to any third-party the model's possessor.
This is the only case that the watermark has forensics value.

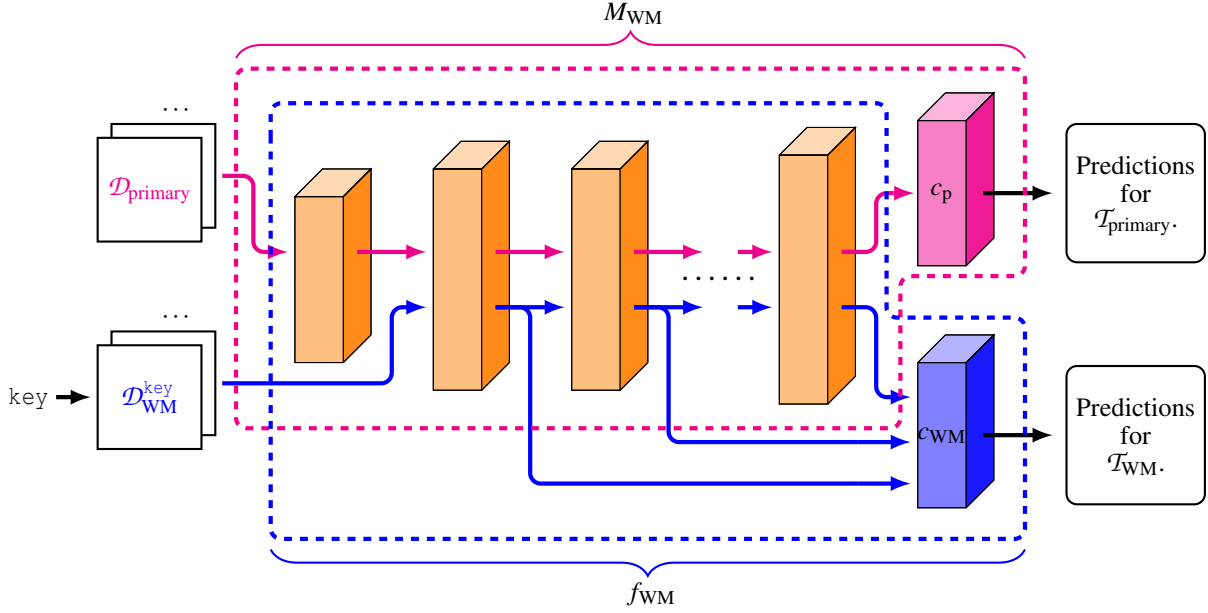
\begin{figure*}[htbp]
\centering
\begin{tikzpicture}[scale=0.92]
\draw [color=black,thick](0+0.2,0+0.2) rectangle (1.5+0.2,-1.5+0.2);
\draw [color=black,thick](0,0) rectangle (1.5,-1.5) [fill=white];
\node at (1.15,0.4) {$\cdots$};
\node at (0.75,-0.75)[color=magenta] {$\mathcal{D}_{\text{primary}}$};
\draw [color=black,thick](0+0.2,0+0.2-3) rectangle (1.5+0.2,-1.5+0.2-3);
\draw [color=black,thick](0,0-3) rectangle (1.5,-1.5-3) [fill=white];
\node at (1.15,0.4-3) {$\cdots$};
\node at (0.75,-0.75-3)[color=blue] {$\mathcal{D}_{\text{WM}}^{\texttt{key}}$};
\node at (-1,-3.75) {\texttt{key}};
\draw [-latex,ultra thick](-0.6,-3.75)--(-0.1,-3.75);
%\draw[decorate,decoration={brace,raise=3pt,amplitude=6pt}] (1.8,-0.55)--(1.8,-3.55);
\draw [rounded corners,color=magenta,ultra thick] (1.8,-0.55)--(2.25,-0.55)--(2.25,-1.65)--(2.55,-1.65);
\draw [-latex,color=magenta,ultra thick] (2.55,-1.65)--(2.75,-1.65);
\draw [rounded corners,color=blue,ultra thick] (1.8,-3.55)--(4.25,-3.55)--(4.25,-2.45)--(4.55,-2.45);
\draw [-latex,color=blue,ultra thick] (4.55,-2.45)--(4.75,-2.45);

\draw [black,fill=orange,fill opacity=0.5] (2.85,-0.85) rectangle (3.55,-3.25);
\draw [black,fill=orange,fill opacity=0.3] (2.85,-0.85)--(2.85+0.4,-0.85+0.4)--(2.85+0.4+0.7,-0.85+0.4)--(3.55,-0.85)--(2.85,-0.85);
\draw [black,fill=orange,fill opacity=0.9] (3.55,-0.85)--(2.85+0.4+0.7,-0.85+0.4)--(2.85+0.4+0.7,-0.85+0.4-2.4)--(3.55,-3.25);
\draw [black] (2.85,-0.85) rectangle (3.55,-3.25);
\draw [black] (2.85,-0.85)--(2.85+0.4,-0.85+0.4)--(2.85+0.4+0.7,-0.85+0.4)--(3.55,-0.85)--(2.85,-0.85);
\draw [black] (3.55,-0.85)--(2.85+0.4+0.7,-0.85+0.4)--(2.85+0.4+0.7,-0.85+0.4-2.4)--(3.55,-3.25);

\draw [-latex,color=magenta,ultra thick] (3.75,-1.65)--(4.75,-1.65);

\draw [black,fill=orange,fill opacity=0.5] (2.85+2,-0.85+0.4) rectangle (3.55+2,-3.25-0.4);
\draw [black,fill=orange,fill opacity=0.3] (2.85+2,-0.85+0.4)--(2.85+0.4+2,-0.85+0.4+0.4)--(2.85+0.4+0.7+2,-0.85+0.4+0.4)--(3.55+2,-0.85+0.4)--(2.85+2,-0.85+0.4);
\draw [black,fill=orange,fill opacity=0.9] (3.55+2,-0.85+0.4)--(2.85+0.4+0.7+2,-0.85+0.4+0.4)--(2.85+0.4+0.7+2,-0.85+0.4-2.4-0.4)--(3.55+2,-3.25-0.4);
\draw [black] (2.85+2,-0.85+0.4) rectangle (3.55+2,-3.25-0.4);
\draw [black] (2.85+2,-0.85+0.4)--(2.85+0.4+2,-0.85+0.4+0.4)--(2.85+0.4+0.7+2,-0.85+0.4+0.4)--(3.55+2,-0.85+0.4)--(2.85+2,-0.85+0.4);
\draw [black] (3.55+2,-0.85+0.4)--(2.85+0.4+0.7+2,-0.85+0.4+0.4)--(2.85+0.4+0.7+2,-0.85+0.4-2.4-0.4)--(3.55+2,-3.25-0.4);

\draw [-latex,color=magenta,ultra thick] (3.75+2,-1.65)--(4.75+2,-1.65);
\draw [-latex,color=blue,ultra thick] (3.75+2,-2.45)--(4.75+2,-2.45);
\draw [rounded corners,color=blue,ultra thick] (5.75,-2.45)--(6.25,-2.45)--(6.25,-6+1)--(11,-6+1);
\draw [-latex,color=blue,ultra thick] (11,-6+1)--(11.75,-6+1);

\draw [black,fill=orange,fill opacity=0.5] (2.85+4,-0.85+0.4) rectangle (3.55+4,-3.25-0.4);
\draw [black,fill=orange,fill opacity=0.3] (2.85+4,-0.85+0.4)--(2.85+0.4+4,-0.85+0.4+0.4)--(2.85+0.4+0.7+4,-0.85+0.4+0.4)--(3.55+4,-0.85+0.4)--(2.85+4,-0.85+0.4);
\draw [black,fill=orange,fill opacity=0.9] (3.55+4,-0.85+0.4)--(2.85+0.4+0.7+4,-0.85+0.4+0.4)--(2.85+0.4+0.7+4,-0.85+0.4-2.4-0.4)--(3.55+4,-3.25-0.4);
\draw [black] (2.85+4,-0.85+0.4) rectangle (3.55+4,-3.25-0.4);
\draw [black] (2.85+4,-0.85+0.4)--(2.85+0.4+4,-0.85+0.4+0.4)--(2.85+0.4+0.7+4,-0.85+0.4+0.4)--(3.55+4,-0.85+0.4)--(2.85+4,-0.85+0.4);
\draw [black] (3.55+4,-0.85+0.4)--(2.85+0.4+0.7+4,-0.85+0.4+0.4)--(2.85+0.4+0.7+4,-0.85+0.4-2.4-0.4)--(3.55+4,-3.25-0.4);

\draw [-latex,color=magenta,ultra thick] (3.75+4,-1.65)--(4.75+4,-1.65);
\draw [-latex,color=blue,ultra thick] (3.75+4,-2.45)--(4.75+4,-2.45);
\node [scale=1.3] at (9,-2.05) {$\cdots\cdots$};
\draw [rounded corners,color=blue,ultra thick] (7.75,-2.45)--(8.25,-2.45)--(8.25,-5.4+1)--(11,-5.4+1);
\draw [-latex,color=blue,ultra thick] (11,-5.4+1)--(11.75,-5.4+1);

\draw [-latex,color=magenta,ultra thick] (3.75+5.5,-1.65)--(4.75+5,-1.65);
\draw [-latex,color=blue,ultra thick] (3.75+5.5,-2.45)--(4.75+5,-2.45);
\draw [black,fill=orange,fill opacity=0.5] (2.85+7,-0.85+0.6) rectangle (3.55+7,-3.25-0.6);
\draw [black,fill=orange,fill opacity=0.3] (2.85+7,-0.85+0.6)--(2.85+0.4+7,-0.85+0.4+0.6)--(2.85+0.4+0.7+7,-0.85+0.4+0.6)--(3.55+7,-0.85+0.6)--(2.85+7,-0.85+0.6);
\draw [black,fill=orange,fill opacity=0.9] (3.55+7,-0.85+0.6)--(2.85+0.4+0.7+7,-0.85+0.4+0.6)--(2.85+0.4+0.7+7,-0.85+0.4-2.4-0.6)--(3.55+7,-3.25-0.6);
\draw [black] (2.85+7,-0.85+0.6) rectangle (3.55+7,-3.25-0.6);
\draw [black] (2.85+7,-0.85+0.6)--(2.85+0.4+7,-0.85+0.4+0.6)--(2.85+0.4+0.7+7,-0.85+0.4+0.6)--(3.55+7,-0.85+0.6)--(2.85+7,-0.85+0.6);
\draw [black] (3.55+7,-0.85+0.6)--(2.85+0.4+0.7+7,-0.85+0.4+0.6)--(2.85+0.4+0.7+7,-0.85+0.4-2.4-0.6)--(3.55+7,-3.25-0.6);

\draw [black,fill=magenta,fill opacity=0.5] (2.85+9,-0.85+1.6-0.5) rectangle (3.55+9,-3.25-0.6+2.5-0.5);
\node at (2.85+0.35+9,-0.3-0.5) {$c_{\text{p}}$};
\draw [black,fill=magenta,fill opacity=0.3] (2.85+9,-0.85+1.6-0.5)--(2.85+0.4+9,-0.85+0.4+1.6-0.5)--(2.85+0.4+0.7+9,-0.85+0.4+1.6-0.5)--(3.55+9,-0.85+1.6-0.5)--(2.85+9,-0.85+1.6-0.5);
\draw [black,fill=magenta,fill opacity=0.9] (3.55+9,-0.85+1.6-0.5)--(2.85+0.4+0.7+9,-0.85+0.4+1.6-0.5)--(2.85+0.4+0.7+9,-0.85+0.4-2.4-0.6+2.5-0.5)--(3.55+9,-3.25-0.6+2.5-0.5);
\draw [black] (2.85+9,-0.85+1.6-0.5) rectangle (3.55+9,-3.25-0.6+2.5-0.5);
\draw [black] (2.85+9,-0.85+1.6-0.5)--(2.85+0.4+9,-0.85+0.4+1.6-0.5)--(2.85+0.4+0.7+9,-0.85+0.4+1.6-0.5)--(3.55+9,-0.85+1.6-0.5)--(2.85+9,-0.85+1.6-0.5);
\draw [black] (3.55+9,-0.85+1.6-0.5)--(2.85+0.4+0.7+9,-0.85+0.4+1.6-0.5)--(2.85+0.4+0.7+9,-0.85+0.4-2.4-0.6+2.5-0.5)--(3.55+9,-3.25-0.6+2.5-0.5);

\draw [rounded corners,color=magenta,ultra thick] (3.75+7,-1.65)--(4.75+6.5,-1.65)--(4.75+6.5,-0.3-0.5)--(5+6.5,-0.3-0.5);
\draw [-latex,color=magenta,ultra thick] (5+6.5,-0.3-0.5)--(5.3+6.5,-0.3-0.5);
\draw [-latex,color=black,ultra thick] (12.8,-0.3-0.5)--(13.8,-0.3-0.5);
\draw [rounded corners,thick] (14,-0.3+1-0.5) rectangle (16,-0.3-1-0.5);
\node at (15,-0.3+0.4-0.5) {Predictions};
\node at (15,-0.3-0.5) {for};
\node at (15,-0.3-0.4-0.5) {$\mathcal{T}_{\text{primary}}$.};
\draw [rounded corners,color=blue,ultra thick] (3.75+7,-2.45)--(4.75+6.5,-2.45)--(4.75+6.5,-4.75+1)--(5+6.5,-4.75+1);
\draw [-latex,color=blue,ultra thick] (11.5,-4.75+1)--(11.75,-4.75+1);

\draw [black,fill=blue,fill opacity=0.5] (2.85+9,-0.85+1.6-5+1) rectangle (3.55+9,-3.25-0.6+2.5-5+1);
\node at (2.85+0.35+9,-5.8+0.5+1) {$c_{\text{WM}}$};
\draw [black,fill=blue,fill opacity=0.3] (2.85+9,-0.85+1.6-5+1)--(2.85+0.4+9,-0.85+0.4+1.6-5+1)--(2.85+0.4+0.7+9,-0.85+0.4+1.6-5+1)--(3.55+9,-0.85+1.6-5+1)--(2.85+9,-0.85+1.6-5+1);
\draw [black,fill=blue,fill opacity=0.9] (3.55+9,-0.85+1.6-5+1)--(2.85+0.4+0.7+9,-0.85+0.4+1.6-5+1)--(2.85+0.4+0.7+9,-0.85+0.4-2.4-0.6+2.5-5+1)--(3.55+9,-3.25-0.6+2.5-5+1);
\draw [black] (2.85+9,-0.85+1.6-5+1) rectangle (3.55+9,-3.25-0.6+2.5-5+1);
\draw [black] (2.85+9,-0.85+1.6-5+1)--(2.85+0.4+9,-0.85+0.4+1.6-5+1)--(2.85+0.4+0.7+9,-0.85+0.4+1.6-5+1)--(3.55+9,-0.85+1.6-5+1)--(2.85+9,-0.85+1.6-5+1);
\draw [black] (3.55+9,-0.85+1.6-5+1)--(2.85+0.4+0.7+9,-0.85+0.4+1.6-5+1)--(2.85+0.4+0.7+9,-0.85+0.4-2.4-0.6+2.5-5+1)--(3.55+9,-3.25-0.6+2.5-5+1);

\draw [-latex,color=black,ultra thick] (12.8,-5.8+0.5+1)--(13.8,-5.8+0.5+1);
\draw [rounded corners,thick] (14,-5.8+1+0.5+1) rectangle (16,-5.8-1+0.5+1);
\node at (15,-5.8+0.4+0.5+1) {Predictions};
\node at (15,-5.8+0.5+1) {for};
\node at (15,-5.8-0.4+0.5+1) {$\mathcal{T}_{\text{WM}}$.};

\draw [rounded corners,dashed,ultra thick,color=magenta] (2,0)--(2,1.5-0.5)--(13.4,1.5-0.5)--(13.4,-2)--(11.6,-2)--(11.6,-4.2)--(2,-4.2)--(2,0);
\draw [decorate,decoration={brace,raise=4pt,amplitude=10pt},thick,color=magenta] (2,1.5-0.5)--(13.4,1.5-0.5);
\node at (7.75,2.3-0.5) {$M_{\text{WM}}$};

\draw [rounded corners,dashed,ultra thick,color=blue] (2.5,0)--(2.5,1-0.5)--(11.4,1-0.5)--(11.4,-3+0.4)--(13.4,-3+0.4)--(13.4,-7.6+0.8+1)--(2.5,-7.6+0.8+1)--(2.5,0);
\draw [decorate,decoration={brace,raise=4pt,amplitude=10pt,mirror},thick,color=blue] (2.5,-7.6+0.8+1)--(13.4,-7.6+0.8+1);
\node at (8,-8.3+0.7+1) {$f_{\text{WM}}$};

\end{tikzpicture}
\caption{Architecture of the MTL-based watermarking scheme. The orange blocks are the backbone, the pink block is the backend for $\mathcal{T}_{\text{primary}}$, the blue block is the classifier for $\mathcal{T}_{\text{WM}}$.}
\label{figure:3.1}
\end{figure*}

The scheme in~\cite{le2020adversarial} is a zero-bit watermarking scheme.
The method proposed by Zhang et al. in~\cite{zhang2018protecting} adopts marked images or noise as the backdoor triggers.
But only a few marks that are easily forgeable were examined.
The protocol of Uchida et al.~\cite{uchida2017embedding} can be enhanced into level III secure against ownership piracy only if an authority is responsible for distributing the secret key, e.g.~\cite{9359144}.
But it lacks covertness and the privacy-preserving property. 

The VAE adopted in~\cite{li2019prove} has to be used conjugately with a secret key that enhances the robustness of the backdoor.
The adversary can collect a set of mistaken samples from one class, slightly disturb them, and claim to have watermarked the neural network.
To claim the ownership of a model watermarked by Adi et al.~\cite{adi2018turning}, the adversary samples its collection of triggers from the mistaken samples, encrypts them with a key, and submits the encrypted pairs.
The perfect security of their scheme depends on the model to perform nearly perfect in the primary task, which is unrealistic in practice.
As for \texttt{DeepSigns}~\cite{darvish2019deepsigns}, one adversary can choose one class and compute the empirical mean of the output of the activation functions (since the outliers are easy to detect) then generate a random matrix as the mask and claim ownership.

The scheme in~\cite{zhu2020secure} is of level III secure against ownership piracy as proved in the original paper.
So is the method in~\cite{li2019persistent} since it is generally hard to guess the actual pattern of the \texttt{Wonder Filter} mask from a space with size $2^{P}$, where $P$ is the number of pixels of the mask.
The scheme by Guan et al. in~\cite{guan2020reversible} is secure but extremely fragile, hence is out of the scope of practical watermarking schemes.

A comprehensive summary of established watermarking schemes judged according to the enumerated security requirements is given in Table \ref{table:1}.

\section{The Proposed Method}
\label{section:3}
\subsection{Motivation}
It is difficult for the backdoor-based or weight-based watermarking methods to formally meet all the proposed security requirements.
Hence, we design a new white-box watermarking method for DNN model protection using multiple task learning.
The watermark embedding is designed as an additional task $\mathcal{T}_{\text{WM}}$.
A classifier for $\mathcal{T}_{\text{WM}}$ is built independent to the backend for $\mathcal{T}_{\text{primary}}$. 
After training and watermark embedding, only the network structure for $\mathcal{T}_{\text{primary}}$ is published. 

Reverse engineering or backdoor detection as~\cite{8835365} cannot find any evidence of the watermark.
Since no trigger is embedded in the published model's backend.   
On the other hand, common FT methods such as fine-tune last layer (FTLL) or re-train last layers (RTLL)~\cite{adi2018turning} that only modifies the backend layers of the model have no impact to our watermark.

Under this formulation, the functionality-preserving property, the security against tuning, the security against watermark detection and privacy-preserving can be formally addressed.
A decently designed $\mathcal{T}_{\text{WM}}$ ensures the security against ownership piracy as well, making the MTL-based watermarking scheme a secure and sound option for model protection.

To better handle the forensic difficulties involving overwritten watermark and key management, we introduce a decentralized consensus protocol to authorize the time stamp embedded with the watermarks. 

\subsection{Overview}
The proposed model consists of the MTL-based watermarking scheme and the decentralized verification protocol. 

\subsubsection{The MLT-based watermarking scheme}
\label{section:3.2.1}
The structure of our watermarking scheme is illustrated in Fig.\ref{figure:3.1}.
The entire network consists of the backbone network and two independent backends: $c_{\text{p}}$ and $c_{\text{WM}}$. 
The published model $M_{\text{WM}}$ is the backbone followed by $c_{\text{p}}$.
While $f_{\text{WM}}$ is the \emph{watermarking branch} for the watermarking task, in which $c_{\text{WM}}$ takes the output of different layers from the backbone as its input.
By having $c_{\text{WM}}$ monitor the outputs of different layers of the backbone network, it is harder for an adversary to design modifications to invalid $c_{\text{WM}}$ completely. 

To produce a watermarked model, a host should:
\begin{enumerate}
\item Generate a collection of $N$ samples
$
\mathcal{D}^{\texttt{key}}_{\text{WM}}=\left\{x_{i},y_{i} \right\}_{i=1}^{N}
$
using a pseudo-random algorithm with \texttt{key} as the random seed.
\item Optimize the entire DNN to jointly minimize the loss on $\mathcal{D}^{\texttt{key}}_{\text{WM}}$ and $\mathcal{D}_{\text{primary}}$.
During the optimization, a series of regularizers are designed to meet the security requirements enumerated in Section \ref{section:2}. 
\item Publishes $M_{\text{WM}}$. 
\end{enumerate}

To prove its ownership over a model $M$ to a third-party:
\begin{enumerate}
\item The host submits $M$, $c_{\text{WM}}$ and $\texttt{key}$. 
\item The third-party generates $\mathcal{D}_{\text{WM}}^{\texttt{key}}$ with $\texttt{key}$ and combines $c_{\text{WM}}$ with $M$'s backbone to build a DNN for $\mathcal{T}_{\text{WM}}$.
\item If the statistical test indicates that $c_{\text{WM}}$ with $M$'s backbone performs well on $\mathcal{D}_{\text{WM}}^{\texttt{key}}$ then the third-party confirms the host's ownership over $M$.
\end{enumerate}

\subsubsection{The decentralized verification protocol}
To enhance the reliability of the ownership protection, it is necessary to use a protocol to authorize the watermark of the model's host.  
Otherwise any adversary who has downloaded $M_{\text{WM}}$ can embed its watermark into it and pirate the model. 

One option is to use an trusted key distribution center or a timing agency, which is in charge of authorizing the time stamp of the  hosts' watermarks.
However, such centralized protocols are vulnerable and expensive.
For this reason we resort to decentralized consensus protocols such as Raft~\cite{ongaro2014search} or PBFT~\cite{castro1999practical}, which were designed to synchronize message within a distributed community.
Under these protocols, one message from a user is responded and recorded by a majority of clients within the community so this message becomes authorized and unforgeable. 

Concretely, a client $s$ under this DNN watermarking protocol is given a pair of public key and private key.
$s$ can publish a watermarked model or claim its ownership over some model by broadcasting:

\textbf{Publishing a model:}
After finishing training a model watermarked with$\texttt{key}$, $s$ obtains $M_{\text{WM}}$ and $c_{\text{WM}}$. 
Then $s$ signs and broadcasts the following message to the entire community:
$$\langle\textbf{Publish:}\texttt{key}\|\texttt{time}\|\texttt{hash}(c_{\text{WM}}) \rangle,$$
where $\|$ denotes string concatenation, $\texttt{time}$ is the time stamp, and $\texttt{hash}$ is a preimage resistant hash function mapping a model into a string and is accessible for all clients.
Other clients within the community verify this message using $s$'s public key, verify that $\texttt{time}$ lies within a recent time window and write this message into their memory. 
Once $s$ is confirmed that the majority of clients has recorded its broadcast (e.g. when $s$ receives a confirmation from the current leader under the Raft protocol), it publishes $M_{\text{WM}}$. 

\textbf{Proving its ownership over a model $M$:}
$s$ signs and broadcasts the following message:
$$\langle\textbf{Claim:}l_{M}\|\texttt{hash}(M)\|l_{c_{\text{WM}}} \rangle,$$
where $l_{M}$ and $l_{c_{\text{WM}}}$ are pointers to $M$ and $c_{\text{WM}}$. 
Upon receiving this request, any client can independently conduct the ownership proof. 
It firstly downloads the model from $l_{M}$ and examines its hash.
Then it downloads $c_{\text{WM}}$ and retrieves the $\textbf{Publish}$ message from $s$ by $\texttt{hash}(c_{\text{WM}})$.
The last steps follow Section.~\ref{section:3.2.1}. 
After finishing the verification, this client can broadcast its result as the proof for $s$'s ownership over the model in $l_{M}$. 

\subsection{Security Analysis of the Watermark Task}
We now elaborate the design of the watermarking task $\mathcal{T}_{\text{WM}}$ and analyze its security.
For simplicity, $\mathcal{T}_{\text{WM}}$ is instantiated as a binary classification task, i.e., the output of the watermarking branch has two channels. 
To generate $\mathcal{D}^\texttt{key}_{\text{WM}}$, $\texttt{key}$ is used as the seed of a pseudo-random generator (e.g., a stream cipher) to generate $\pi^{\texttt{key}}$, a sequence of $N$ different integers from the range $[0,\cdots,2^{m}-1]$, and a binary string $\texttt{l}^{\texttt{key}}$ of length $N$, where $m=3 \lceil \log_{2}(N)\rceil$. 

For each type of data space $\mathcal{X}$, a deterministic and injective function is adopted to map each interger in $\pi^{\texttt{key}}$ into an element in $\mathcal{X}$.
For example, when $\mathcal{X}$ is the image domain, the mapping could be the QRcode encoder.
When $\mathcal{X}$ is the sequence of words in English, the mapping could map an integer $n$ into the $n$-th word of the dictionary.\footnote{We suggest not to use function that encodes integers into terms that are similar to data in $\mathcal{T}_{\text{primary}}$, especially to data of the same class. 
This increase the difficulty for $c_{\text{WM}}$ to achieve perfect classification.} 
Without loss of generality, let $\pi^{\texttt{key}}[i]$ denotes the mapped data from the $i$-th integer in $\pi^{\texttt{key}}$. 
Both the pseudo-random generator and the functions that map integers into specialized data space should be accessible for all clients within the intellectual property protection community. 
Now we set:
$$
\mathcal{D}^{\texttt{key}}_{\text{WM}}=\left\{(\pi^{\texttt{key}}_{m}[i],\texttt{l}^{\texttt{key}}[i]) \right\}_{i=1}^{N},
$$
where $\texttt{l}^{\texttt{key}}[i]$ is the $i$-th bit of $\texttt{l}$. 
We now merge the security requirements raised in Section \ref{section:2} into this framework.

\subsubsection{The correctness}
To verify the ownership of a model $M$ to a host with $\texttt{key}$ given $c_{\text{WM}}$, the process $\texttt{verify}$ operates as Algo. \ref{algorithm:2}. 
\begin{algorithm}[htbp] 
\caption{$\texttt{verify}(\cdot,\cdot | c_{\text{WM}},\gamma)$}  
\label{algorithm:2}  
\begin{algorithmic}[1]
\REQUIRE $M$, $\texttt{key}$.
\ENSURE The verification of $M$'s ownership.
\STATE Build the watermarking branch $f$ from $M$ and $c_{\text{WM}}$;
\STATE Generate $\mathcal{D}^{\texttt{key}}_{\text{WM}}$ from $\texttt{key}$;   
\IF {$f$ correctly classifies at least $\gamma\cdot N$ terms within $\mathcal{D}^{\texttt{key}}_{\text{WM}}$}
\RETURN 1.
\ELSE 
\RETURN 0.
\ENDIF
\end{algorithmic}  
\end{algorithm} 

If $M=M_{\text{WM}}$ then $M$ has been trained to minimize the binary classification loss on $\mathcal{T}_{\text{WM}}$, hence the test is likely to succeed in Algo.~\ref{algorithm:2}, this justifies the requirement from~\eqref{equation:1}.
For an arbitrary $\texttt{key}'\neq \texttt{key}$, the induced watermark training data $\mathcal{D}^{\texttt{key}'}_{\text{WM}}$ can hardly be similar to $\mathcal{D}^{\texttt{key}}_{\text{WM}}$. 
To formulate this intuition, consider the event where $\mathcal{D}^{\texttt{key}'}_{\text{WM}}$ shares $q\cdot N$ terms with $\mathcal{D}^{\texttt{key}}_{\text{WM}}$, $q \in (0,1)$. 
With a pseudorandom generator, it is computationally impossible to distinguish $\pi^{\texttt{key}}$ from an sequence of $N$ randomly selected intergers.
The same argument holds for $\texttt{l}^{\texttt{key}}$ and a random binary string of length $N$. 
Therefore the probability of this event can be upper bounded by:
$$
%\begin{aligned}
\binom{N}{qN}\cdot r^{qN}\cdot\left(1-r \right)^{(1-q)N}
%&\leq \left[\frac{\frac{n}{2}\cdots(\frac{(1-q)\cdot n}{2}+1)}{\frac{q\cdot n}{2}\cdots 1} \right]^{2}(\frac{r}{1-r})^{q\cdot n}\\
\leq \left[\left(1+(1-q)N\right) \left(\frac{r}{1-r}\right) \right]^{qN},
%\end{aligned}
$$
where $r=\frac{N}{2^{m+1}}$. 
For an arbitrary $q$, let 
$r< \frac{1}{2+(1-q)N}$
then the probability that $\mathcal{D}^{\texttt{key}'}_{\text{WM}}$ overlaps with $\mathcal{D}^{\texttt{key}}_{\text{WM}}$ with a portion of $q$ declines exponentially. 

For numbers not appeared in $\pi^{\texttt{key}}$, the watermarking branch is expected to output a random guess. 
Therefore if $q$ is smaller than a threshold $\tau$ then $\mathcal{D}^{\texttt{key}'}_{\text{WM}}$ can hardly pass the statistical test in Algo.\ref{algorithm:2} with $n$ big enough. 
So let 
$$m\geq \log_{2}\left[ 2N\left(2+(1-\tau)N \right)\right]$$ 
and $n$ be large enough would make an effective collision in the watermark dataset almost impossible. 
For simplicity, setting $m=3\cdot \lceil \log_{2}(N)\rceil \geq \log_{2}(N^{3})$ is sufficient.

In cases $M_{\text{WM}}$ is replaced by an arbitrary model whose backbone structure happens to be consistent with $c_{\text{WM}}$, the output of the watermarking branch remains a random guess.
This justifies the second requirement for correct verification~\eqref{equation:2}. 

To select the threshold $\gamma$, assume that the random guess strategy achieves an average accuracy of at most $p=0.5+\alpha$, where $\alpha\geq 0$ is a bias term which is assumed to decline with the growth of $n$. 
The verification process returns 1 iff the watermark classifier achieves binary classification of accuracy no less than $\gamma$. 
The demand for security is that by randomly guessing, the probability that an adversary passes the test declines exponentially with $n$. 
Let $X$ denotes the number of correct guessing with average accuracy $p$, an adversary suceeds only if $X\geq \gamma\cdot N$.
By the Chernoff theorem:
$$\text{Pr}\left\{X\geq \gamma\cdot N \right\}\leq \left(\frac{1-p+p\cdot\text{e}^{\lambda}}{\text{e}^{\gamma\cdot \lambda}} \right)^{N},$$
where $\lambda$ is an arbitrary nonnegative number. 
If $\gamma$ is larger than $p$ by a constant independent of $N$ then $\left(\frac{1-p+p\cdot\text{e}^{\lambda}}{\text{e}^{\gamma\cdot \lambda}} \right)$ is less than 1 with proper $\lambda$, reducing the probability of successful attack into negligibility.  

\subsubsection{The functionality-preserving regularizer}
Denote the trainable parameters of the DNN model by $\textbf{w}$. 
The optimization target for $\mathcal{T}_{\text{primary}}$ takes the form:
\begin{equation}
\label{equation:8}
\mathcal{L}_{0}(\textbf{w},\mathcal{D}_{\text{primary}})=\sum_{(x,y)\in\mathcal{D}_{\text{primary}}}l\left(M^{\textbf{w}}_{\text{WM}}(x),y\right)+\lambda_{0}\cdot u(\textbf{w}) ,
\end{equation}
where $l$ is the loss defined by $\mathcal{T}_{\text{primary}}$ and $u(\cdot)$ is a regularizer reflecting the prior knowledge on $\textbf{w}$. 
The normal training process computes the empirical loss in \eqref{equation:8} by stochastically sampling batches and adopting gradient-based optimizers. 

The proposed watermarking task adds an extra data dependent term to the loss function:
\begin{equation}
\label{equation:9}
\begin{aligned}
\mathcal{L}(\textbf{w},\mathcal{D}_{\text{primary}},\mathcal{D}_{\text{WM}})&=\mathcal{L}_{0}(\textbf{w},\mathcal{D}_{\text{primary}})\\
&+\lambda\cdot\sum_{(x,y)\in\mathcal{D}_{\text{WM}}}l_{\text{WM}}\left(f^{\textbf{w}}_{\text{WM}}(x),y\right),
\end{aligned}
\end{equation}
where $l_{\text{WM}}$ is the cross entropy loss for binary classification.
We omitted the dependency of $\mathcal{D}_{\text{WM}}$ on $\texttt{key}$ in this section for conciseness. 

To train multiple tasks, we can minimize the loss function for multiple tasks \eqref{equation:9} directly or train the watermarking task and the primary task alternatively~\cite{mtl}. 
Since $\mathcal{D}_{\text{WM}}$ is much smaller than $\mathcal{D}_{\text{primary}}$, it is possible that $\mathcal{T}_{\text{WM}}$ does not properly converge when being learned simultaneously with $\mathcal{T}_{\text{primary}}$. 
Hence we first optimize $\textbf{w}$ according to the loss on the primary task \eqref{equation:8} to obtain $\textbf{w}_{0}$:
$$
\textbf{w}_{0}=\arg\min_{\textbf{w}}\left\{\mathcal{L}_{0}(\textbf{w},\mathcal{D}_{\text{primary}}) \right\}.
$$
 
Next, instead of directly optimizing the network w.r.t. \eqref{equation:9}, the following loss function is minimized: 
\begin{equation}
\label{equation:10}
\begin{aligned}
\mathcal{L}_{1}(\textbf{w},\mathcal{D}_{\text{primary}},\mathcal{D}_{\text{WM}})=&\sum_{(x,y)\in\mathcal{D}_{\text{WM}}}l_{\text{WM}}(f^{\textbf{w}}_{\text{WM}}(x),y)\\
&+\lambda_{1}\cdot R_{\text{func}}(\textbf{w}),\\
\end{aligned}
\end{equation}
where
\begin{equation}
\label{equation:r1}
R_{\text{func}}(\textbf{w})=\|\textbf{w}-\textbf{w}_{0}\|_{2}^{2}.
\end{equation}
By introducing the regularizer $R_{\text{func}}$ in \eqref{equation:r1}, $\textbf{w}$ is confined in the neighbour of $\textbf{w}_{0}$. 
Given this constraint and the continuity of $M_{\text{WM}}$ as a function of $\textbf{w}$, we can expect the functionality-preserving property defined in \eqref{equation:3}. 
Then the weaker version of functionality-preserving \eqref{equation:4} is tractable as well. 

\subsubsection{The tuning regularizer}
To be secure against adversary's tuning, it is sufficient to make $c_{\text{WM}}$ robust against tuning by the definition in~\eqref{equation:6}. 
Although $\mathcal{D}_{\text{adversary}}$ is unknown to the host, we assume that $\mathcal{D}_{\text{adversary}}$ shares a similar distribution as $\mathcal{D}_{\text{primary}}$.
Otherwise the stolen model would not have the state-of-the-art performance on the adversary's task. 
To simulate the influence of tuning, a subset of $\mathcal{D}_{\text{primary}}$ is firstly sampled as an estimation of $\mathcal{D}_{\text{adversary}}$:
$
\mathcal{D}'_{\text{primary}}\xleftarrow{\text{sample}} \mathcal{D}_{\text{primary}}
$.
Let $\textbf{w}$ be the current configuration of the model's parameter. 
Tuning is usually tantanmount to minimizing the empirical loss on $\mathcal{D}'_{\text{primary}}$ by starting from $\textbf{w}$, which results in an updated parameter:
$
\textbf{w}^{\text{t}}\xleftarrow[\mathcal{D}'_{\text{primary}}]{\text{tune}}\textbf{w}
$.
In practice, $\textbf{w}^{\text{t}}$ is obtained by replacing $\mathcal{D}_{\text{primary}}$ in \eqref{equation:8} by $\mathcal{D}'_{\text{primary}}$ and conducting a few rounds of gradient descents from $\textbf{w}$. 

To achieve the security against tuning defined in~\eqref{equation:6}, it is sufficient that the parameter $\textbf{w}$ satisfies:
\begin{equation}
\label{equation:12}
\begin{aligned}
\forall \mathcal{D}'_{\text{primary}}&\xleftarrow{\text{sample}} \mathcal{D}_{\text{primary}}, \textbf{w}^{\text{t}}\xleftarrow[\mathcal{D}'_{\text{primary}}]{\text{tune}}\textbf{w},\forall (x,y) \in \mathcal{D}_{\text{WM}},\\ 
&f^{\textbf{w}^{\text{t}}}_{\text{WM}}(x)=y. \\
\end{aligned}
\end{equation}
The condition \eqref{equation:12}, Algo.\ref{algorithm:1} together with the assumption that $\mathcal{D}_{\text{adversary}}$ is similar to $\mathcal{D}_{\text{primary}}$ imply \eqref{equation:6}. 

To exert the constraint in \eqref{equation:12} to the training process, we design a new regularizer as follows:
\begin{equation}
\label{equation:r2}
R_{\text{DA}}(\textbf{w})=\sum_{
\begin{scriptsize}
\begin{aligned}
&\mathcal{D}'_{\text{primary}}\xleftarrow{\text{sample}} \mathcal{D}_{\text{primary}},\\
&\textbf{w}^{\text{t}}\xleftarrow[\mathcal{D}'_{\text{primary}}]{\text{tune}} \textbf{w},(x,y)\in\mathcal{D}_{\text{WM}}\\
\end{aligned}
\end{scriptsize}
}l_{\text{W}}\left(f^{\textbf{w}^{\text{t}}}_{\text{WM}}(x),y\right).
\end{equation} 
Then the loss to be optimized is updated from \eqref{equation:10} to:
\begin{equation}
\label{equation:14}
\mathcal{L}_{2}(\textbf{w},\mathcal{D}_{\text{primary}},\mathcal{D}_{\text{WM}})=\mathcal{L}_{1}(\textbf{w},\mathcal{D}_{\text{primary}},\mathcal{D}_{\text{WM}})+\lambda_{2}\cdot R_{\text{DA}}(\textbf{w}).
\end{equation}

$R_{\text{DA}}$ defined by~\eqref{equation:r2} can be understood as one kind of \emph{data augmentation} for $\mathcal{T}_{\text{WM}}$.
Data augmentation aims to improve the model's robustness against some specific perturbation in the input.
This is done by proactively adding such perturbation to the training data. 
According to~\cite{shorten2019survey}, data augmentation can be formulated as an additional regularizer:
\begin{equation}
\label{equation:15}
\sum_{(x,y)\in\mathcal{D},x'\xleftarrow{\text{perturb}}x}l\left(f^{\textbf{w}}(x'),y\right).
\end{equation}

Unlike in the ordinary data domain of $\mathcal{T}_{\text{primary}}$, it is hard to explicitly define augmentation for $\mathcal{T}_{\text{WM}}$ against tuning.
However, a regularizer with the form of \eqref{equation:15} can be derived from \eqref{equation:r2} by interchanging the order of summation so the perturbation takes the form:
$$x'\in\left[f^{\textbf{w}}_{\text{WM}} \right]^{-1}\left( f^{\textbf{w}^{\text{t}}}_{\text{WM}}\left(x\right) \right)\xleftarrow{\text{perturb}}x.$$

\subsubsection{Security against watermark detection}
Consider the extreme case where $\lambda_{1}\rightarrow \infty$.
Under this configuration, the parameters of $M_{\text{WM}}$ are frozen and only the parameters in $c_{\text{WM}}$ are tuned. 
Therefore $M_{\text{WM}}$ is exactly the same as $M_{\text{clean}}$ and it seems that we have not insert any information into the model. 
However, by broadcasting the designed message, the host can still prove that it has obtained the white-box access to the model at an early time, which fact is enough for ownership verification. 
This justifies the security against watermark detection by the definition of \eqref{equation:detection}, where $\lambda_{1}$ casts the role of $\theta$.

\subsubsection{Privacy-preserving}
Recall the definition of privacy-preserving in Section \ref{section:2.3.4}.
We prove that, under certain configurations, the proposed watermarking method is privacy-preserving.
\begin{theorem}
Let $c_{\text{WM}}$ take the form of a linear classifier whose input dimensionality is $L$. 
If $N\leq (L+1)$ then the watermarking scheme is secure against assignment detection.
\end{theorem}
\begin{proof}
The VC-dimension of a linear classifier with $L$ channels is $(L+1)$.
Therefore for $N\leq (L+1)$ inputs with arbitrary binary labels, there exists one $c_{\text{WM}}$ that can almost always perfectly classify them. 
Given $M$ and an arbitrary $\texttt{key}'$, it is possible forge $c_{\text{WM}}'$ such that $c_{\text{WM}}'$ with $M$'s backbone performs perfectly on $\mathcal{D}_{\text{WM}}^{\texttt{key}'}$. 
We only have to plug the parameters of $M$ into~\eqref{equation:14}, set $\lambda_{1}\rightarrow\infty$, $\lambda_{2}=0$ and minimize the loss. 
This step ends up with a watermarked model $M_{\text{WM}}=M$ and an evidence, $c_{\text{WM}}'$, for $\texttt{key}'$. 
Hence for the experiment defined in Algo.~\ref{exp:1}, an adversary cannot identify the host's key since evidence for both options are equally plausible. 
The adversary can only conduct a random guess, whose probability of success is $\frac{1}{2}$. 
\end{proof}
This theorem indicates that, the MTL-based watermarking scheme can protect the host's privacy. 
Moreover, given $N$, it is crucial to increase the input dimensionality of $c_{\text{WM}}$ or using a sophiscated structure for $c_{\text{WM}}$ to increase its VC-dimensionality.

\subsubsection{Security against watermark overwriting}
It is possible to meet the definition of the security against watermark overwriting in \eqref{equation:7} by adding the perturbation of embedding other secret keys into $R_{\text{DA}}$.
But this requires building other classifier structures and is expensive even for the host.
For an adversary with insufficient training data, it is common to freeze the weights in the backbone layers as in transfer learning~\cite{pan2009survey}, hence~\eqref{equation:7} is satisfied. 
For general cases, an adversary would not disturb the backbone of the DNN too much for the sake of its functionality on the primary task. 
Hence we expect the watermarking branch to remain valid after overwriting. 

We leave the examination of the security against watermark overwriting as an empirical study.

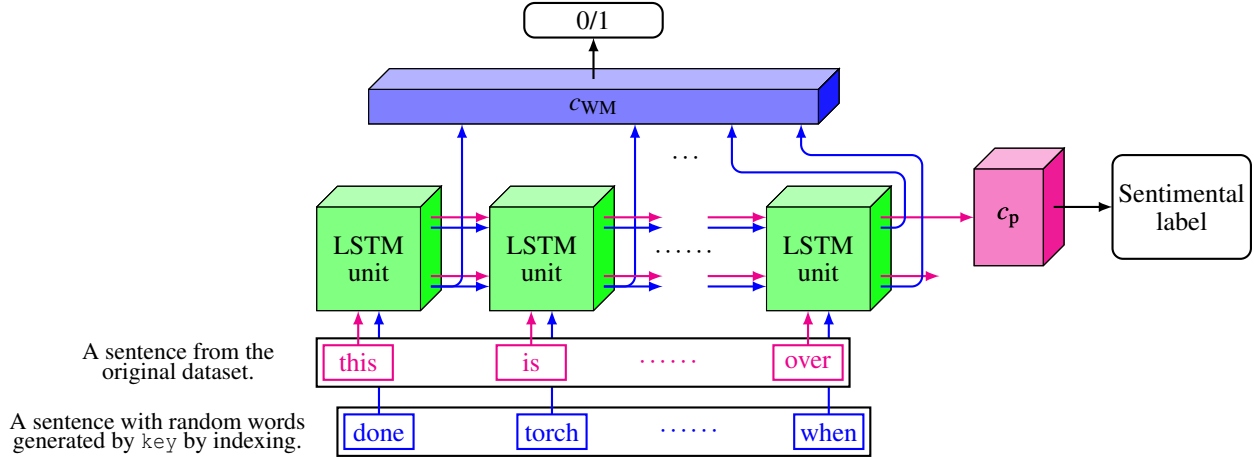
\begin{figure*}[htbp]
\centering
\begin{tikzpicture}[scale=0.92]
\draw [fill=green,fill opacity=0.5] (0,0) rectangle (1.5,1.5);
\draw [fill=green,fill opacity=0.3] (0,1.5)--(0.3,1.8)--(1.8,1.8)--(1.5,1.5);
\draw [fill=green,fill opacity=0.9] (1.5,1.5)--(1.8,1.8)--(1.8,0.3)--(1.5,0);
\draw [black] (0,0) rectangle (1.5,1.5);
\draw [black] (0,1.5)--(0.3,1.8)--(1.8,1.8)--(1.5,1.5);
\draw [black] (1.5,1.5)--(1.8,1.8)--(1.8,0.3)--(1.5,0);

\draw [fill=green,fill opacity=0.5] (0+2.5,0) rectangle (1.5+2.5,1.5);
\draw [fill=green,fill opacity=0.3] (0+2.5,1.5)--(0.3+2.5,1.8)--(1.8+2.5,1.8)--(1.5+2.5,1.5);
\draw [fill=green,fill opacity=0.9] (1.5+2.5,1.5)--(1.8+2.5,1.8)--(1.8+2.5,0.3)--(1.5+2.5,0);
\draw [black] (0+2.5,0) rectangle (1.5+2.5,1.5);
\draw [black] (0+2.5,1.5)--(0.3+2.5,1.8)--(1.8+2.5,1.8)--(1.5+2.5,1.5);
\draw [black] (1.5+2.5,1.5)--(1.8+2.5,1.8)--(1.8+2.5,0.3)--(1.5+2.5,0);

\draw [fill=green,fill opacity=0.5] (0+6.5,0) rectangle (1.5+6.5,1.5);
\draw [fill=green,fill opacity=0.3] (0+6.5,1.5)--(0.3+6.5,1.8)--(1.8+6.5,1.8)--(1.5+6.5,1.5);
\draw [fill=green,fill opacity=0.9] (1.5+6.5,1.5)--(1.8+6.5,1.8)--(1.8+6.5,0.3)--(1.5+6.5,0);
\draw [black] (0+6.5,0) rectangle (1.5+6.5,1.5);
\draw [black] (0+6.5,1.5)--(0.3+6.5,1.8)--(1.8+6.5,1.8)--(1.5+6.5,1.5);
\draw [black] (1.5+6.5,1.5)--(1.8+6.5,1.8)--(1.8+6.5,0.3)--(1.5+6.5,0);

\draw [blue,thick] (0.1+0.3,-1-1) rectangle (1.1+0.3,-0.5-1);
\node [blue] at (0.9,-1.75) {done};
\draw [blue,thick] (0.1+2.5+0.3,-1-1) rectangle (1.1+2.5+0.3,-0.5-1);
\node [blue] at (0.9+2.5,-1.75) {torch};
\draw [blue,thick] (0.1+6.5+0.3,-1-1) rectangle (1.1+6.5+0.3,-0.5-1);
\node [blue] at (0.9+6.5,-1.75) {when};
\node [blue,thick] at (0.6+4.5+0.3,-0.75-1) {$\cdots\cdots$};
\draw [thick] (0+0.3,-1.1-1) rectangle (1.1+6.5+0.1+0.3,-0.4-1);
\draw [blue,-latex,thick] (0.6+0.3,-1.5)--(0.6+0.3,0);
\draw [blue,-latex,thick] (0.6+2.5+0.3,-1.5)--(0.6+2.5+0.3,0);
\draw [blue,-latex,thick] (0.6+6.5+0.3,-1.5)--(0.6+6.5+0.3,0);
\node [scale=0.9] at (-2.3,-1.75+0.16) {A sentence with random words};
\node [scale=0.9] at (-2.3,-1.75-0.16) {generated by $\texttt{key}$ by indexing.};

\draw [fill=white,thick] (0,-1.1) rectangle (1.1+6.5+0.1,-0.4);
\draw [magenta,thick] (0.1,-1) rectangle (1.1,-0.5);
\node [magenta] at (0.6,-0.75) {this};
\draw [magenta,thick] (0.1+2.5,-1) rectangle (1.1+2.5,-0.5);
\node [magenta] at (0.6+2.5,-0.75) {is};
\draw [magenta,thick] (0.1+6.5,-1) rectangle (1.1+6.5,-0.5);
\node [magenta] at (0.6+6.5,-0.75) {over};
\node [magenta] at (0.6+4.5,-0.75) {$\cdots\cdots$};
\draw [magenta,-latex,thick] (0.6,-0.5)--(0.6,0);
\draw [magenta,-latex,thick] (0.6+2.5,-0.5)--(0.6+2.5,0);
\draw [magenta,-latex,thick] (0.6+6.5,-0.5)--(0.6+6.5,0);
\node [scale=0.9] at (-2,-0.75+0.16) {A sentence from the};
\node [scale=0.9] at (-2,-0.75-0.16) {original dataset.};

\draw [magenta,-latex,thick] (1.65,1.35)--(2.5,1.35);
\draw [blue,-latex,thick] (1.65,1.2)--(2.5,1.2);
\draw [magenta,-latex,thick] (1.65,0.5)--(2.5,0.5);
\draw [blue,-latex,thick] (1.65,0.35)--(2.5,0.35);
\draw [rounded corners, blue,-latex,thick] (1.65,0.35)--(2.1,0.35)--(2.1,3-0.3);
\draw [rounded corners, blue,-latex,thick] (1.65+2.5,0.35)--(2.1+2.5,0.35)--(2.1+2.5,3-0.3);

\draw [magenta,-latex,thick] (1.65+2.5,1.35)--(2.5+2.5,1.35);
\draw [blue,-latex,thick] (1.65+2.5,1.2)--(2.5+2.5,1.2);
\draw [magenta,-latex,thick] (1.65+2.5,0.5)--(2.5+2.5,0.5);
\draw [blue,-latex,thick] (1.65+2.5,0.35)--(2.5+2.5,0.35);

\draw [magenta,-latex,thick] (1.65+4,1.35)--(2.5+4,1.35);
\draw [blue,-latex,thick] (1.65+4,1.2)--(2.5+4,1.2);
\draw [magenta,-latex,thick] (1.65+4,0.5)--(2.5+4,0.5);
\draw [blue,-latex,thick] (1.65+4,0.35)--(2.5+4,0.35);
\node at (5.325,0.85) {$\cdots\cdots$};
\node at (0.75,0.75+0.2) {LSTM};
\node at (0.75,0.75-0.2) {unit};
\node at (0.75+2.5,0.75+0.2) {LSTM};
\node at (0.75+2.5,0.75-0.2) {unit};
\node at (0.75+6.5,0.75+0.2) {LSTM};
\node at (0.75+6.5,0.75-0.2) {unit};
\draw [magenta,-latex,thick] (1.65+6.5,1.35)--(9.5,1.35);
\draw [rounded corners,blue,-latex,thick] (1.65+6.5,1.2)--(2+6.5,1.2)--(8.5,2)--(6,2)--(6,3-0.3);
\draw [magenta,-latex,thick] (1.65+6.5,0.5)--(2.5+6.5,0.5);
\draw [rounded corners,blue,-latex,thick] (1.65+6.5,0.35)--(2.25+6.5,0.35)--(8.75,2.25)--(7,2.25)--(7,3-0.3);
\node at (5.35,2.7-0.5) {$\cdots$};

\draw [fill=magenta,fill opacity=0.5] (9.5,0.65) rectangle (10.5,2.05);
\node at (10,1.35) {$c_{\text{p}}$};
\draw [fill=magenta,fill opacity=0.3] (9.5,2.05)--(9.5+0.3,2.05+0.3)--(10.5+0.3,2.05+0.3)--(10.5,2.05);
\draw [fill=magenta,fill opacity=0.9] (10.5,2.05)--(10.5+0.3,2.05+0.3)--(10.5+0.3,0.65+0.3)--(10.5,0.65);
\draw [black] (9.5,0.65) rectangle (10.5,2.05);
\node at (10,1.35) {$c_{\text{p}}$};
\draw [black] (9.5,2.05)--(9.5+0.3,2.05+0.3)--(10.5+0.3,2.05+0.3)--(10.5,2.05);
\draw [black] (10.5,2.05)--(10.5+0.3,2.05+0.3)--(10.5+0.3,0.65+0.3)--(10.5,0.65);
\draw [-latex,thick] (10.65,1.5)--(11.5,1.5);
\draw [rounded corners,thick] (11.5,0.75) rectangle (13.5,2.25);
\node at (12.5,1.5+0.2) {Sentimental};
\node at (12.5,1.5-0.2) {label};

\draw [fill=blue,fill opacity=0.5] (0.75,3-0.3) rectangle (0.75+6.5,3.5-0.3);
\draw [fill=blue,fill opacity=0.3] (0.75,3.5-0.3)--(0.75+0.3,3.8-0.3)--(0.75+6.5+0.3,3.8-0.3)--(0.75+6.5,3.5-0.3);
\draw [fill=blue,fill opacity=0.9] (0.75+6.5,3.5-0.3)--(0.75+6.5+0.3,3.5+0.3-0.3)--(0.75+6.5+0.3,3.3-0.3)--(0.75+6.5,3-0.3);
\draw [black] (0.75,3-0.3) rectangle (0.75+6.5,3.5-0.3);
\draw [black] (0.75,3.5-0.3)--(0.75+0.3,3.8-0.3)--(0.75+6.5+0.3,3.8-0.3)--(0.75+6.5,3.5-0.3);
\draw [black] (0.75+6.5,3.5-0.3)--(0.75+6.5+0.3,3.5+0.3-0.3)--(0.75+6.5+0.3,3.3-0.3)--(0.75+6.5,3-0.3);
\node at (4,3.25-0.3) {$c_{\text{WM}}$};
\draw [-latex,thick] (4,4.15-0.8)--(4,4.75-0.8);
\draw [thick,rounded corners] (3,4.75-0.8)rectangle(5,5.25-0.8);
\node at (4,5-0.8) {0/1};

\end{tikzpicture}
\caption{The network architecture for sentimental analysis.}
\label{figure:nlp}
\end{figure*}

\subsubsection{Security against ownership piracy}
Recall that in ownership piracy, the adversary is not allowed to train its own watermark classifier. 
Instead, it can only forge a key given a model $M_{\text{WM}}$ and a legal $c_{\text{WM}}$, this is possible if the adversary has participated in the proof for some other client.    
Now the adversary is to find a new key $\texttt{key}_{\text{adv}}$ such that $\mathcal{D}^{\texttt{key}_{\text{adv}}}_{\text{WM}}$ can pass the statistical test defined by the watermarking branch $M_{\text{WM}}$ and $c_{\text{WM}}$. 
Although it is easy to find a set of $N$ intergers with half of them classified as $0$ and half $1$ by querying the watermarking branch as an oracle, it is hard to restore a legal $\texttt{key}_{\text{adv}}$ from this set. 
The protocol should adopt a \emph{stream cipher secure against key recovery attack}~\cite{rudskoy2010zero}, which, by definition, blocks this sort of ownership piracy and makes the proposed watermarking scheme of level III secure against ownership piracy. 
If $c_{\text{WM}}$ is kept secret then the ownership piracy is impossible.
Afterall, ownership piracy is invalid when an authorized time stamp is avilable. 

\subsection{Analysis of the Verification Protocol}
We now conduct the security analysis to the consensus protocal and solve the redeclaration dilemma.

To pirate a model under this protocol, an adversary must submit a legal $\texttt{key}$ and the hash of a $c_{\text{WM}}$.
If the adversary does not have a legal $c_{\text{WM}}$ then this attack is impossible since the preimage resistance of $\texttt{hash}$ implies that the adversary cannot forge such a watermark classifier afterwards.
So this broadcast is invalid. 
If the adversary has managed to build a legal $c_{\text{WM}}$, compute its hash, but has not obtained the target model then the verification can hardly succeed since the output of $c_{\text{WM}}$ with the backbone of an unknown network on the watermark dataset is random guessing.
The final case is that the adversary has obtained the target model, conducted the watermark overwriting and redeclared the ownership. 
Recall that the model is published only if its host has successfully broadcast its $\textbf{Publish}$ message and notarized its $\texttt{time}$.
Hence the overwriting dilemma can be solved by comparing the time stamp inside contradictive broadcasts. 

As an adaptive attack, one adversary participating in the proof of a host's ownership over a model $M$ obtains the corresponding $\texttt{key}$ and $c_{\text{WM}}$, with which it can erase weight-based watermarks~\cite{uchida2017embedding,9359144}.
Embedding information into the outputs of the network rather than its weights makes the MTL-based watermark harder to erase. 
The adversary has to identify the decision boundary from $c_{\text{WM}}$ and tune $M$ so samples drawn from $\texttt{key}$ violates this boundary. 
This attack risks the model's performance on the primary task, requires huge amont of data and computation resources and is beyond the competence of a model thief. 

The remaining security risks are within the cryptological components and beyond the scope of our discussion.
%\footnote{For example, $\texttt{hash}$ might fail to be a preimage resistant hash function, or an adversary party might dominate the majority of all clients} 

\section{Experiments and Discussions}
\subsection{Experiment Setup}
To illustrate the flexibility of the proposed watermarking model, we considered four primary tasks: image classification (IC), malware classification (MC), image semantic segmentation (SS) and sentimental analysis (SA) for English. 
We selected four datasets for image classification, one dataset for malware classification, two datasets for semantic segmentation and two datasets for sentimental classification. 
The descriptions of these datasets and the corresponding DNN structures are listed in Table~\ref{table:2}.
ResNet~\cite{he2016deep} is a classical model for image processing.
For the VirusShare dataset, we compiled a collection of 26,000 malware into images and adopted ResNet as the classifier~\cite{chu2020visualization}.
Cascade mask RCNN (CMRCNN)~\cite{cai2018cascade} is a network architecture specialized for semantic segmentation.
Glove~\cite{pennington2014glove} is a pre-trained word embedding that maps English words into numerical vectors, while bidirectional long short-term memory (Bi-LSTM)~\cite{huang2015bidirectional} is commonly used to analyze natural languages.

\begin{table}[htb]
\caption{Datasets and their DNN structures.}
\begin{center}
\begin{tabular}{c|c|c}
\toprule[1.5pt]
\textbf{Dataset} & \textbf{Description} & \textbf{DNN structure} \\
\midrule[1pt]
MNIST~\cite{deng2012mnist} & IC, 10 classes & ResNet-18\\
\midrule
\tabincell{c}{Fashion-\\MNIST~\cite{xiao2017/online}} & IC, 10 classes & ResNet-18\\
\midrule
CIFAR-10~\cite{krizhevsky2009learning} & IC, 10 classes & ResNet-18\\
\midrule
CIFAR-100~\cite{krizhevsky2009learning} & IC, 100 classes& ResNet-18\\
\midrule
VirusShare~\cite{vs} & MC, 10 classes & ResNet-18\\
\midrule
\tabincell{c}{Penn-Fudan\\-Pedestrian~\cite{wang2007object}} & SS, 2 classes & \tabincell{c}{ResNet-50+\\CMRCNN}\\
\midrule
VOC~\cite{voc2012} & SS, 20 classes & \tabincell{c}{ResNet-50+\\CMRCNN}\\
\midrule
IMDb~\cite{maas2011learning} & SA, 2 classes & Glove+Bi-LSTM\\
\midrule
SST~\cite{socher2013recursive} & SA, 5 classes & Glove+Bi-LSTM\\
\bottomrule[1pt]
\end{tabular}
\label{table:2}
\end{center}
\end{table}
\begin{table*}[hbt]
\caption{Ablation study on regularizer configuration. Each entry contains the four metrics in Section~\ref{section:4.2}. Semantic segmentation tasks were measured by average precision and these two models would not converge without $R_{\text{func}}$. The optimal/second optimal configuration for each dataset and each metric are highlighted/underlined.}
\begin{center}
\begin{tabular}{c|c|c|c|c|c}
\toprule[1.5pt]
\multirow{2}{*}{\textbf{Dataset}} & \multirow{2}{*}{\tabincell{c}{$M_{\text{clean}}$\textbf{'s} \\\textbf{performance}}} & \multicolumn{4}{c}{\textbf{Regularizer configuration}}\\
\cline{3-6}
 & & No regularizers. & $R_{\text{func}}$ & $R_{\text{DA}}$ & $R_{\text{func}}$ and $R_{\text{DA}}$\\
\midrule[1pt]
MNIST & 99.6\% & \tabincell{c}{98.7\%,75.5\%,\\85.0\%,1.3\%}  & \tabincell{c}{\textbf{99.5\%},81.5\%,\\ 85.5\%,0.7\%}  & \tabincell{c}{99.3\%,\underline{88.0\%},\\\underline{92.0\%},\underline{2.0\%}} & \tabincell{c}{\textbf{99.5\%},\textbf{95.5\%},\\\textbf{92.5\%},\textbf{2.3\%}}\\
\midrule
\tabincell{c}{Fashion-\\MNIST} & 93.3\% & \tabincell{c}{92.0\%,85.0\%,\\60.5\%,9.6\%} &  \tabincell{c}{\underline{92.8\%},92.0\%,\\ 74.5\%,11.6\%} & \tabincell{c}{91.4\%,\textbf{96.5\%},\\\underline{85.5\%},\underline{54.6\%}} & \tabincell{c}{\textbf{93.1\%},\underline{95.5\%},\\\textbf{86.0\%},\textbf{54.9\%}}\\
\midrule
CIFAR-10 & 91.5\% & \tabincell{c}{88.3\%,91.5\%,\\74.5\%,19.8\%} & \tabincell{c}{\textbf{90.8\%},88.5\%,\\79.5\%,21.0\%} & \tabincell{c}{\underline{88.8\%},\textbf{96.0\%},\\ \textbf{95.0\%},\textbf{56.0\%}} & \tabincell{c}{\underline{88.8\%},\underline{92.5\%},\\ \underline{91.5\%},\underline{53.0\%}} \\
\midrule
CIFAR-100 & 67.7\% & \tabincell{c}{59.9\%,90.5\%,\\88.0\%,23.6\%} & \tabincell{c}{\textbf{65.4\%},90.0\%,\\87.0\%,23.3\%} & \tabincell{c}{58.7\%,\textbf{99.0\%},\\ \textbf{98.0\%},\textbf{35.0\%}}  & \tabincell{c}{\underline{63.8\%},\underline{92.5\%},\\\underline{97.0\%},\underline{33.3\%}}\\
\midrule
VirusShare & 97.4\% & \tabincell{c}{97.0\%,65.0\%,\\88.0\%,6.4\%} & \tabincell{c}{\underline{97.2\%},81.5\%,\\86.5\%,6.5\%} & \tabincell{c}{96.8\%,\underline{99.5\%},\\\textbf{100\%},\underline{9.4\%}} & \tabincell{c}{\textbf{97.3\%},\textbf{100\%},\\\textbf{100\%},\textbf{19.5\%}}\\
\midrule
\tabincell{c}{Penn-Fudan-\\Pedestrian} & 0.79 & --- & \tabincell{c}{\underline{0.79},\underline{90.0\%},\\\underline{54.5\%},\underline{0.70}} & --- & \tabincell{c}{\textbf{0.78},\textbf{100\%},\\\textbf{100\%},\textbf{0.78}} \\
\midrule
VOC & 0.69 & --- & \tabincell{c}{\underline{0.67},\underline{74.0\%},\\\underline{98.0\%},\underline{0.65}} & --- & \tabincell{c}{\textbf{0.69},\textbf{100\%},\\\textbf{100\%},\textbf{0.68}} \\
\midrule
IMDb & 85.0\% & \tabincell{c}{67.3\%,66.8\%,\\83.5\%,12.0\%} & \tabincell{c}{\textbf{85.0\%},66.0\%,\\86.3\%,12.2\%}  & \tabincell{c}{69.2\%,\textbf{81.3\%},\\\underline{88.3\%},\underline{29.5\%}} & \tabincell{c}{\textbf{85.0\%},\underline{80.0\%},\\\textbf{90.8\%},\textbf{30.5\%}} \\
\midrule
SST & 75.4\% & \tabincell{c}{71\%,77.3\%,\\95.8\%,12.5\%} & \tabincell{c}{\textbf{75.4\%},62.5\%,\\95.0\%,13.0\%} & \tabincell{c}{70.8\%,\textbf{90.5\%},\\\underline{98.3\%},\underline{29.4\%}} & \tabincell{c}{\textbf{75.4\%},\underline{86.8\%},\\\textbf{99.0\%},\textbf{31.9\%}} \\
\bottomrule[1pt]
\end{tabular}
\label{table:3}
\end{center}

\end{table*}
\begin{table}[htbp]
\caption{Fluctuation of the accuracy of the host's watermarking branch.}
\begin{center}
\begin{tabular}{c|m{1cm}<{\centering}|m{1cm}<{\centering}|m{1cm}<{\centering}|m{1cm}<{\centering}}
\toprule[1.5pt]
\multirow{2}{*}{\textbf{Dataset}} & \multicolumn{4}{c}{\textbf{Number of overwriting epochs}}\\
\cline{2-5}
 & 50 & 150 &250& 350  \\
\midrule[1pt]
MNIST & 1.0\% &1.5\% &1.5\% &2.0\%  \\
\midrule
\tabincell{c}{Fashion-\\MNIST} & 2.0\% &2.5\%  &2.5\%  &2.5\% \\
\midrule
CIFAR-10 & 4.5\% & 4.5\% & 4.5\% & 4.5\% \\
\midrule
CIFAR-100 & 0.0\% & 0.5\%  & 0.9\% & 0.9\%\\
\midrule
VirusShare & 0.0\% & 0.5\% & 0.5\% & 0.5\%\\
\midrule
\tabincell{c}{Penn-Fudan-\\Pedestrian} & 0.5\% & 1.0\% & 1.0\% & 1.0\% \\
\midrule
VOC &  1.3\% & 2.0\% & 2.1\% & 2.1\%\\
\midrule
IMDb & 3.0\% & 3.0\% & 3.0\% & 3.0\% \\
\midrule
SST & 2.5\% & 3.0\% & 3.0\% & 2.5\%\\
\bottomrule[1pt]
\end{tabular}
\label{table:4}
\end{center}
\end{table}

For the first seven image datasets, $c_{\text{WM}}$ was a two-layer perceptron  that took the outputs of the first three layers from the ResNet as input. 
QRcode was adopted to generate $\mathcal{D}_{\text{WM}}^{\texttt{key}}$.
For the NLP datasets, the network took the structure in Fig.~\ref{figure:nlp}.

\begin{figure}[htbp]
\centering
\includegraphics[width=6cm]{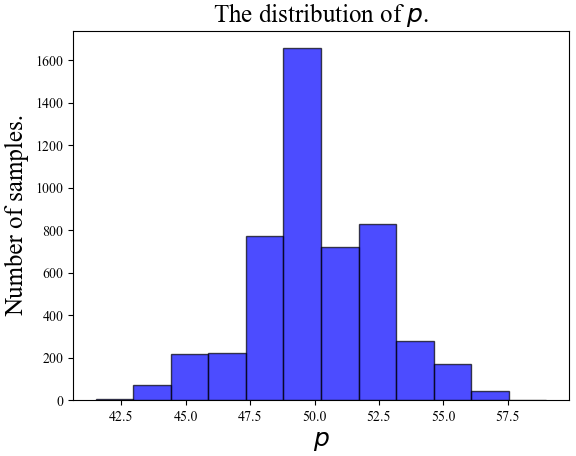}
\caption{The empirical distribution of $p$.}
\label{figure:p}
\end{figure}

Throughout the experiments we set $N=600$. 
To set the verification threshold $\gamma$ in Algo.~\ref{algorithm:2}, we test the classification accuracy of $f_{\text{WM}}$ across nine datasets over 5,000 $\mathcal{D}_{\text{WM}}$s different from the host's. 
The result is visualized in Fig.~\ref{figure:p}, from which we observed that almost all cases $p$ fell in $[0.425,0.575]$.
We selected $\gamma=0.7$ so the probability of success piracy is less than $2.69\times 10^{-8}$ with $\lambda=0.34$ in the Chernoff bound.

We conducted three tuning attacks: FT, NP, FP, and the overwriting attack to the proposed watermarking framework.

\subsection{Ablation Study}
\label{section:4.2}
To examine the efficacy of $R_{\text{func}}$ and $R_{\text{DA}}$, we compared the performance of the model under different combinations of two regularizers.
We are interested in four metrics: (1) the performance of $M_{\text{WM}}$ on $\mathcal{T}_{\text{primary}}$, (2) the performance of $f_{\text{WM}}$ on $\mathcal{T}_{\text{WM}}$ after FT, (3) the performance of $f_{\text{WM}}$ on $\mathcal{T}_{\text{WM}}$ after FP, and (4) the decline of the performance of $M_{WM}$ on $\mathcal{T}_{\text{primary}}$ when NP made $f_{\text{WM}}$'s accuracy on $\mathcal{T}_{\text{WM}}$ lower than $\gamma$. 
The first metric reflects the decline of a model's performance after being watermarked.
The second and the third metrics measure the watermark's robustness against an adversary's tuning.
The last metric reflects the decrease of the model's utility when an adversary is determined to erase the watermark using NP. 
The model for each dataset was trained by minimizing the MTL loss defined by~\eqref{equation:14}, where we adopted FT, NP and FP for tuning and chose the optimal $\lambda_{1}$ and $\lambda_{2}$ by grid search.
Then we attacked each model by FT with a smaller learning rate, FP \cite{liu2018fine} and NP. 
The results are collected in Table~\ref{table:3}. 

We observe that by using $R_{\text{func}}$ and $R_{\text{DA}}$, it is possible to preserve the watermarked model's performance on the primary task and that on the watermarking task simultaneously. 
Therefore we suggest that whenever possible, the two regularizers should be incorporated in training the model. 

\subsection{Watermark Detection}
\begin{figure}[htbp]
\centering
\includegraphics[width=5.7cm]{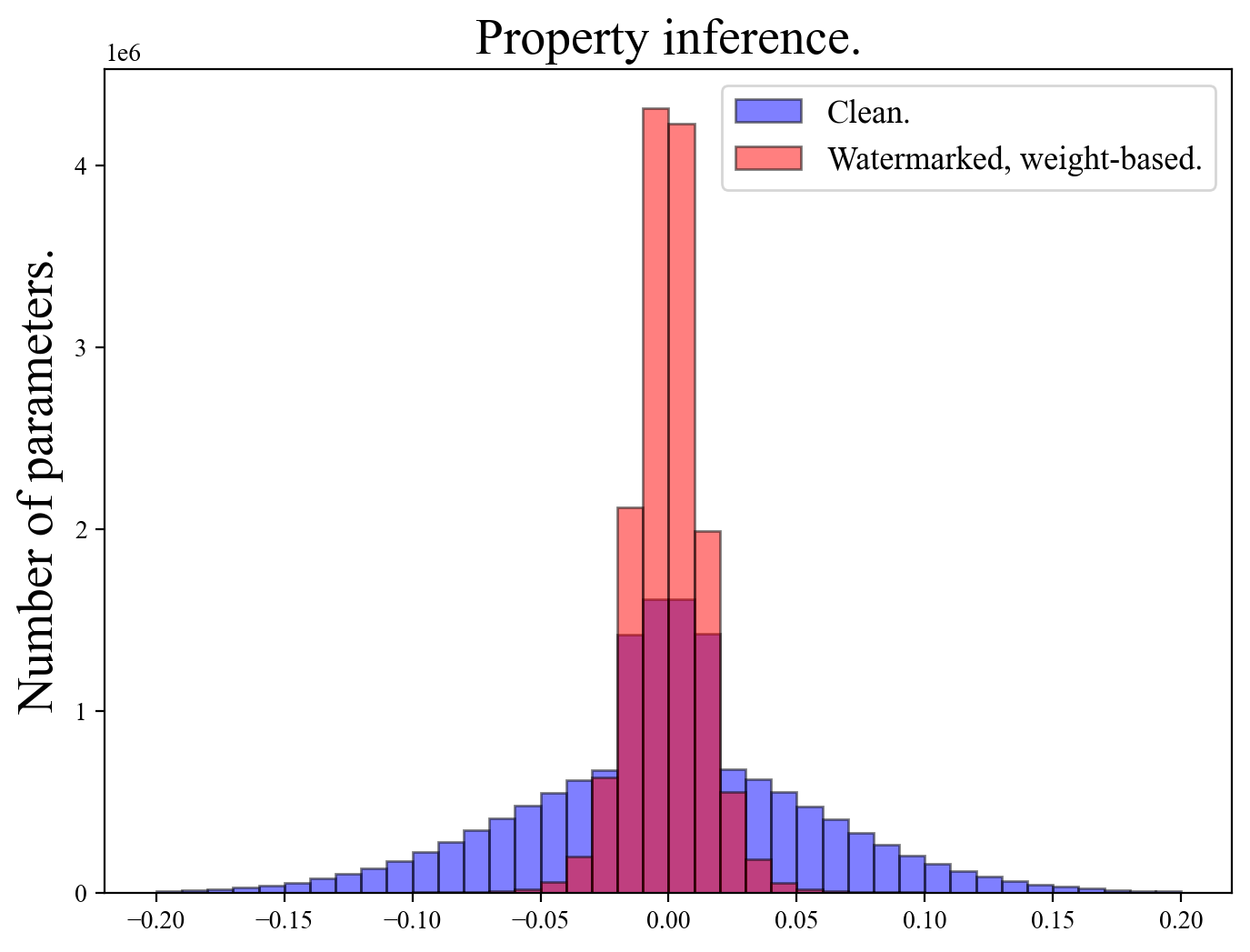}
\caption{The difference between $M_{\text{clean}}$ and a weight-based watermarked model~\cite{darvish2019deepsigns}.}
\label{figure:wd.1}
\end{figure}

\begin{figure}[htbp]
\centering
\includegraphics[width=5.7cm]{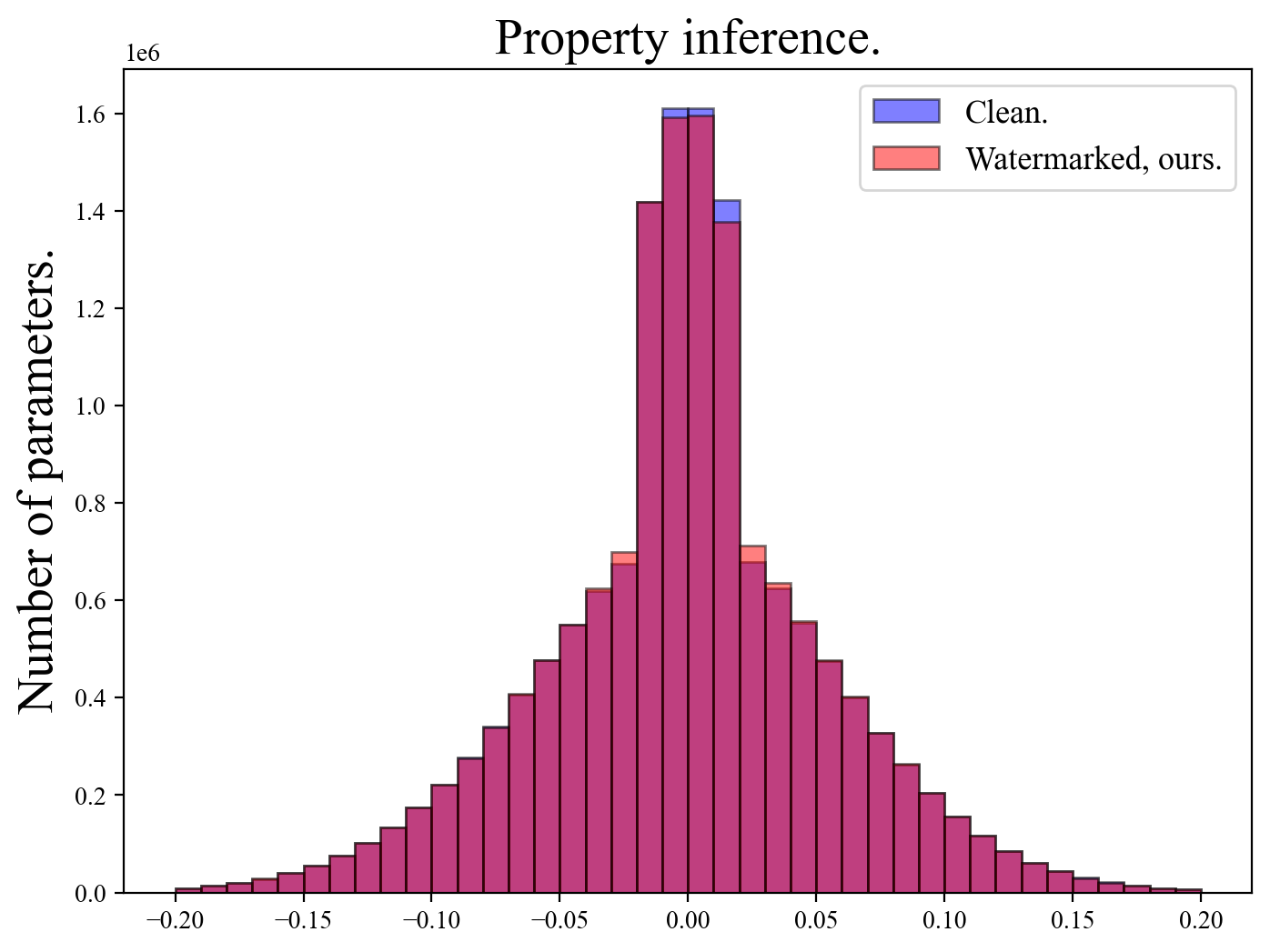}
\caption{The difference between $M_{\text{WM}}$ and $M_{\text{clean}}$.}
\label{figure:wd.2}
\end{figure}

As an illustration of the security against watermark detection, we illustrated the property inference attack~\cite{ganju2018property}. 
The distributions of the parameters of a clean model, a model watermarked by our method and one weight-based method~\cite{darvish2019deepsigns} for CIFAR-10 are visualized in Fig.~\ref{figure:wd.1} and Fig.~\ref{figure:wd.2}. 
In which we adopted $\lambda_{1}=0.05$. 
Unlike the weight-based watermarking method analyzed in~\cite{ganju2018property}, our method did not result in a significant difference between the distributions of parameters of the two models. 
Hence an adversary can hardly distinguish a model watermarked by the MTL-based method from a clean one. 

\begin{table*}[htb]
\caption{The comparision between our method and~\cite{li2019persistent,zhu2020secure} with respect to: (1) the model's performance on the primary task, (2) the accuracy of the watermarking task/backdoor after FP, (3) the decline of the model's accuracy on the primary task when NP erase the watermark. The optimal method for each dataset with respect to each metric is highlighted.}
\begin{center}
\begin{tabular}{c|c|c|c|c|c|c|c|c|c}
\toprule[1.5pt]
\multirow{2}{*}{\textbf{Dataset}} & \multicolumn{3}{c|}{Ours, $R_{\text{func}}$ and $R_{\text{DA}}$} & \multicolumn{3}{c|}{Li et al.~\cite{li2019persistent}} & \multicolumn{3}{c}{Zhu et al.~\cite{zhu2020secure}}\\
\cline{2-10}
 & Primary & FP & NP & Primary & FP & NP& Primary & FP & NP \\
\midrule[1pt]
MNIST & \textbf{99.5\%} &  \textbf{92.5\%} &  \textbf{2.2\%} & 99.0\% & 14.5\% & 0.9\% & 98.8\% & 7.0\% & 1.4\% \\
\midrule
Fashion-MNIST &   \textbf{93.1\%} &  \textbf{86.0\%} &  \textbf{54.7\%} & 92.5\% & 13.5\% & 17.5\% & 91.8\% & 11.3\% & 5.8\% \\
\midrule
CIFAR-10 &  \textbf{88.8\%} &  \textbf{91.5\%} &  \textbf{50.3\%} & 88.5\% & 14.5\% & 13.6\% & 85.0\% & 10.0\% & 17.1\% \\
\midrule
CIFAR-100 &  \textbf{63.8\%} &  \textbf{97.0\%} &  \textbf{29.4\%} & 63.6\% & 1.2\% & 5.5\% & 65.7\% & 0.8\% & 0.9\%\\
\midrule
VirusShare &  \textbf{97.3\%} &  \textbf{100\%} &  \textbf{9.6\%} &  95.1\% & 8.8\% & 1.5\% & 96.3\% & 9.5\% & 1.1\% \\
\bottomrule[1pt]
\end{tabular}
\label{table:5}
\end{center}
\end{table*}

\subsection{The Overwriting Attack}
After adopting both regularizers, we performed overwriting attack to models for all nine tasks, where each model was embedded different keys.
In all cases the adversary's watermark could be successfully embedded into the model, as what we have predicted. 
The metric is the fluctuation of the watermarking branch on the watermarking task after overwriting, as indicated by \eqref{equation:7}.
We recorded the fluctuation for the accuracy of the watermarking branch with the overwriting epoches.
The results are collected in Table~\ref{table:4}. 

The impact of watermark overwriting is uniformly bounded by 4.5\% in our settings. 
And the accuracy of the watermarking branch remained above the threshold $\gamma=0.7$.
Combined with Table~\ref{table:3}, we conclude that the MTL-based watermarking method is secure against watermark overwriting.

\subsection{Comparision and Discussion}
We implemented the watermarking methods in~\cite{zhu2020secure} and~\cite{li2019persistent}, which are both backdoor-based method of level III secure against ownership piracy.
We randomly generated 600 trigger samples for~\cite{zhu2020secure} and assigned them with proper labels.
For~\cite{li2019persistent}, we randomly selected $\texttt{Wonder Filter}$ patterns and exerted them onto 600 randomly sampled images. 

As a comparison, we list the performance of their watermarked models on the primary task, the verification accuracy of their backdoors after FP, whose damage to backdoors is larger than FT, and the decline of the performance of the watermarked models when NP was adopted to invalid the backdoors (when the accuracy of the backdoor triggers is under 15\%) in Table.~\ref{table:5}.
We used the ResNet-18 DNN for all experiments and conducted experiments for the image classifications, since otherwise the backdoor is undefined.

We observe that for all metrics, our method achieved the optimal performance, this is due to:
\begin{enumerate}
\item Extra regularizers are adopted to explicitly meet the security requirements.
\item The MTL-based watermark does not incorporate backdoors into the model, so adversarial modifications such as FP, which are designed to eliminate backdoor, can hardly reduce our watermark.
\item The MTL-based watermark relies on an extra module, $c_{\text{WM}}$, as a verifier.
As an adversary cannot tamper with this module, universal tunings as NP have less impact. 
\end{enumerate}

Apart from these metrics, our proposal is better than other backdoor-based DNN watermarking methods since:
\begin{enumerate}
\item Backdoor-based watermarking methods are not privacy-preserving.
\item So far, backdoor-based watermarking methods can only be applied to image classification DNNs.
This fact challenges the generality of backdoor-based watermark.
\item It is hard to design adaptive backdoor against specific screening algorithms.
However, the MTL-based watermark can easily adapt to new tuning operators. 
This can be done by incorporating such tuning operator into $R_{\text{DA}}$.
\end{enumerate}

\section{Conclusion}
This paper presents a MTL-based DNN watermarking model for ownership verification. 
We summarize the basic security requirements for DNN watermark formally and raise the privacy concern.
Then we propose to embed watermark as an additional task parallel to the primary task. 
The proposed scheme explicitly meets various security requirements by using corresponding regularizers.
Those regularizers and the design of the watermarking task grant the MTL-based DNN watermarking scheme tractable security. 
With a decentralized consensus protocol, the entire framework is secure against all possible attacks. 

We are looking forward to using cryptological protocols such as zero-knowledge proof to improve the ownership verification process so it is possible to use one secret key for multiple notarizations.

%-------------------------------------------------------------------------------
\section*{Acknowledgments}
%-------------------------------------------------------------------------------

This work receives support from anonymous reviewers.

%-------------------------------------------------------------------------------
\section*{Availability}
%-------------------------------------------------------------------------------

Materials of this paper, including source code and part of the dataset, are available at \url{http://github.com/a_new_account/xxx}.

%-------------------------------------------------------------------------------
\bibliographystyle{plain}
\bibliography{WM.bib}

%%%%%%%%%%%%%%%%%%%%%%%%%%%%%%%%%%%%%%%%%%%%%%%%%%%%%%%%%%%%%%%%%%%%%%%%%%%%%%%%
\end{document}